\documentclass[sigconf, nonacm, screen]{acmart}

\newcommand\vldbpagestyle{plain} 

\usepackage[a-2b]{pdfx}
\usepackage{xcolor}
\usepackage{colortbl}
\usepackage{xspace}
\usepackage{subcaption}
\usepackage{amsmath}
\usepackage{listings}
\usepackage{algorithm}
\usepackage{algpseudocodex}

\usepackage{pifont}
\usepackage{eso-pic}
\usepackage{dsfont}
\usepackage{amsthm}

\usepackage{multirow}
\usepackage{multicol}
\usepackage{lipsum}

\usepackage{tcolorbox}
\tcbuselibrary{breakable}

\usepackage{tikz}
\usetikzlibrary{shapes.geometric, arrows}

\usepackage{hhline}

\usepackage{graphicx}
\usepackage[inline,shortlabels]{enumitem}
\usepackage{textcomp}

\usepackage{hyperref}
\usepackage{cleveref}

\usepackage{float}

\usepackage{makecell}
\usepackage{array}

\newcommand{\sparagraph}[1]{\vspace{1mm}\noindent {\bf #1.}}
\newcommand{\thesystem}{\textsc{AutoSLO}\xspace}

\definecolor{darkgreen}{RGB}{0, 100, 0}
\definecolor{experimentblue}{RGB}{127, 159, 186}
\definecolor{experimentgreen}{RGB}{127, 186, 130}
\definecolor{experimentred}{RGB}{186, 138, 127}
\definecolor{darkorange}{RGB}{240,131,15}

\newcommand{\yes}{{\textcolor{experimentgreen}{\ding{51}}}}

\newcommand{\no}{{\textcolor{experimentred}{\ding{55}}}}

\newcounter{rowctr}

\newtcolorbox[auto counter,number within=subsection]{sql}[1][]{
  boxrule=0pt, 
  boxsep=3pt, 
  left=32pt, 
  right=2pt, 
  top=2pt, 
  bottom=2pt, 
  box align=center, 
  halign=left, 
  valign=center, 
  colbacktitle=experimentblue,
  fonttitle=\scshape,
  #1
}

\newtcolorbox[auto counter]{findings}[1]{
  breakable, boxrule=0pt, boxsep=3pt,left=0pt,right=0pt,top=0pt,bottom=0pt, colbacktitle=experimentgreen, title= \thesubsection{} #1: Findings,
}

\newtcolorbox[auto counter]{defbox}[2]{
  breakable, boxrule=0pt, boxsep=3pt,left=0pt,right=0pt,top=0pt,bottom=0pt, colbacktitle=experimentblue, title=Definition \thetcbcounter: #2, label=#1,
}

\newtcolorbox[auto counter,number within=subsection]{simplebox}[0]{
  breakable, boxrule=0pt, boxsep=3pt,left=0pt,right=0pt,top=0pt,bottom=0pt,
}

\theoremstyle{definition}

\newcommand{\baseline}[1]{\textsf{#1}}
\newcommand{\component}[1]{\textsf{#1}}

\begin{document}
\title{\thesystem: Practical Latency SLOs on Cloud Data Warehouses -- Extended Version}

\author{Markos Markakis}
\affiliation{%
  \institution{MIT CSAIL}
  \city{Cambridge}
  \state{MA, USA}
  \country{}
}
\email{markakis@mit.edu}

\author{Tim Kraska}
\affiliation{%
  \institution{MIT CSAIL}
  \city{Cambridge}
  \state{MA, USA}
  \country{}
}
\email{kraska@mit.edu}

\renewcommand{\shortauthors}{Markos Markakis and Tim Kraska}

\begin{abstract}

Modern cloud data warehouses decouple compute from storage, making it easy for organizations to access the same underlying data with multiple compute clusters. This flexibility is often used for performance isolation among diverse workloads, so that each workload meets its latency service-level objective (SLO) more reliably. For example, interactive dashboards, ad hoc analysis, and batch jobs can each run on separate clusters. However, this dedicated-cluster approach requires each compute cluster to be continuously scaled to adapt to workload evolution, with over-provisioning wasting resources and under-provisioning risking SLO violations.

We present \thesystem, a latency-SLO-aware workload management framework for multi-cluster cloud data warehouses. \thesystem operates across three timescales through three key components. First, a periodic \component{Policy Tuner} plans proactive cluster scaling actions and tunes configuration parameters, using simulations of history-derived workload forecasts. Second, an SLO-aware reactive \component{Autoscaler} adjusts the active cluster set when 
recent workload behavior deviates from the forecast. Third, an online \component{Query Router} reacts to live load when placing each query, using a concurrency-aware latency predictor to avoid SLO violations.

On realistic Redbench workloads, \thesystem
successfully meets latency SLOs of varying strictness, reducing cost by a mean of $26.4\%$ compared to the per-scenario next-best baseline. 
Component-level evaluations show that the \component{Query Router}
and \component{Autoscaler} respectively reduce SLO violation
rates by a mean of $47.8\%$ and $93.7\%$, relative to their corresponding alternatives.
Finally, we show that the \component{Policy Tuner} can reduce the SLO violation rate by a mean of $44.6\%$ using a single day of workload history, and that each component is efficient given its intended operating
timescale.

\end{abstract}

\maketitle

\pagestyle{\vldbpagestyle}

\section{Introduction}\label{sec:introduction}

\sparagraph{The Problem} Modern cloud data warehouses such as Amazon Redshift Serverless~\cite{aws-redshift} and Snowflake~\cite{snowflake} decouple compute from storage, enabling multiple compute clusters to access the same underlying data. For example, multiple Amazon Redshift Serverless workgroups can access the same data through Datashares~\cite{aws-redshift-datashares}, while Snowflake natively supports multi-cluster warehouses~\cite{snowflake-multicluster}. 

Customers have embraced this feature~\cite{renen2024why}, because it enables coarse-grained performance isolation by allowing the separation of workloads with different latency service-level objectives (SLOs). For example, interactive dashboards may need to respond within a few seconds, nightly ETL workloads may need to finish before business hours, and data-science workloads may tolerate some additional latency as long as costs are kept bounded. 

However, even if workloads are each assigned to a separate cluster, current systems do not allow SLOs to be specified directly. This means that administrators have to experiment with the cluster's settings (e.g. number of nodes or qualitative price-performance hint~\cite{nathan2024intelligent}) until the desired latency SLO is met. Even after such tuning, there is no control over what happens as queries \emph{within} each workload may interfere with one another. As a recent work by the Amazon Redshift team notes, \emph{``large ad-hoc queries can have a severe negative impact on the overall performance of the cluster, since they can occupy compute resources and cause cache thrashing.''}~\cite{nathan2024intelligent}

In this work, we take a different approach: we treat latency SLOs as first-class inputs and use the mechanisms exposed by cloud data warehouses to meet them cost-efficiently. The goal is not cross-workload performance isolation for its own sake, but rather meeting the SLO \emph{of each individual query}, while minimizing infrastructure cost. This requires deciding which clusters to use, when to scale them, and where to route each incoming query.

\sparagraph{What Success Looks Like} 
If each cluster could be spun up instantaneously and immediately offer its full performance, the task at hand would be simple: the system could wait until demand appears, create exactly the resources needed, and route queries accordingly. However, it takes time to provision new resources, and additional time for caches and other internal state to warm up. As a result, to cost-efficiently meet latency SLOs, we must manage cluster configuration across multiple timescales, as summarized in Table~\ref{tab:desiderata}:

\begin{itemize}[leftmargin=*]

    \item \textbf{Plan ($\approx24$ hrs):}
    Enterprise workloads often exhibit recurring patterns~\cite{renen2024why}. By extracting and anticipating these patterns, the system can make long-term cluster-management and configuration decisions proactively, such as when to spin up or tear down clusters and how to tune the policies used during execution. Because these decisions are made offline and amortized over longer periods, one can afford to evaluate a richer set of alternatives.
    
    \item \textbf{Adjust ($\approx5$ mins):} Forecasts are imperfect, and workloads can evolve. The system must be able to respond by spinning up or tearing down clusters to compensate for unexpected bursts, lulls, or shifts in workload mix. Unlike planning, adjustment operates under tighter time constraints and therefore makes more localized decisions within the policy chosen by the planner. 
    
    \item \textbf{React ($\approx1$ s):} Finally, each arriving query must be assigned to one of the currently active clusters. Since the goal is no longer workload performance isolation, but rather meeting query-level latency SLOs cost-efficiently, this decision must account for the query's own latency SLO, its  impact on already-running queries, and the infrastructure cost implied by the routing decision.
\end{itemize}

These behaviors are complementary. Planning reduces dependence on slow reactive scaling by preparing resources before expected demand arrives. Adjustment handles forecast error and unexpected workload changes. Reaction handles the per-query effects of concurrency and contention, which remain present even under a well-chosen cluster configuration. Together, they allow a system to leverage multiple clusters to meet query-level latency SLOs without requiring rigid workload-to-cluster assignments.

\begin{table}[]

\setlength{\tabcolsep}{1.75pt}

\caption{No prior work exhibits all key desired behaviors.}
\label{tab:desiderata}

\begin{tabular}{l|c|c|c}
\toprule
\textbf{Key} & \textbf{Plan}& \textbf{Adjust}& \textbf{React}  \\
\textbf{Desired} & resources based & resources based&to concurrent \\
 \textbf{Behavior}& on history & on demand& live load \\
\hline
\textbf{Timescale} &$\approx24$ hrs&$\approx5$ mins&$\approx1$ s\\

    \hhline{=|=|=|=}
    
    ResTune&\yes&\no&\no\\
    \cite{zhang2021restune}&\textit{Replay-based}&&\\
    &\textit{knob tuning}&&\\
    &\textit{on replica}&&\\
    \hline
    
    WiseDB &\yes&\no&\no\\
    \cite{marcus2016wisedb}&\textit{Tree model}&&\\
    &\textit{for fixed per-}&&\\
    &\textit{template routing}&&\\

    \hline
    
    Das et al.&\no&\yes&\no\\
    \cite{das2016automated}&&\textit{Container sizing}&\\
    &&\textit{w/ token bucket}&\\
    \hline
    
    SLAOrches-&\no&\yes&\no\\
    trator&&\textit{Utilization- or}&\\
    \cite{ortiz2018slaorchestrator, ortiz2015changing, ortiz2016perfenforce}&&\textit{simulation-based}&\\
    &&\textit{pool resizing}&\\
    \hline
    
    BRAD&\no&\yes&\no\\
    \cite{yu2024blueprinting, kraska2023check}&&\textit{Blueprint}&\\
    &&\textit{beam search}&\\
    \hline
    
    Auto-WLM&\no&\yes&\no\\
    \cite{saxena2023auto}&&\textit{Queueing-based}&\\
    &&\textit{horizontal scaling}&\\
    \hline
    
    RAIS&\yes&\yes&\no\\
    \cite{nathan2024intelligent}&\textit{Multi-forecast}&\textit{Queueing-based}&\\
    &\textit{parameter tuning}&\textit{horizontal scaling}&\\
   
    \hhline{=|=|=|=}
    \textbf{\thesystem} &\yes&\yes&\yes\\
    &\textit{Multi-forecast}&\textit{Simulation-based}&\textit{Concurrency-}\\
    &\textit{spinup planning}&\textit{best-cluster}&\textit{aware routing}\\
    &\textit{\& tuning}&\textit{spinup}&\\

    \hline
    
    \textbf{Component}&\component{Policy Tuner}&\component{Autoscaler}&\component{Query Router}\\
    \hline
    
    \textbf{Details in...}&Section~\ref{sec:policy-tuner}&Section~\ref{sec:autoscaler}&Section~\ref{sec:query-router}\\
\bottomrule
\end{tabular}

\end{table}

\sparagraph{How Prior Work Falls Short} Several automatic workload management mechanisms have been proposed in recent years. However, as shown in Table~\ref{tab:desiderata}, none exhibits all three \textbf{key desired behaviors}. 

Some systems can only \textbf{plan} for a known or forecasted workload, using replays~\cite{zhang2021restune} or modeling~\cite{marcus2016wisedb}. However, planning alone is insufficient: forecasts are imperfect, workload mixes can shift, and live query-level contention can affect SLO adherence and must be managed.
Other systems can only \textbf{adjust} resources online in response to demand, using various forms of short-term performance modeling/assumptions under different available resources~\cite{das2016automated, ortiz2018slaorchestrator, ortiz2015changing, ortiz2016perfenforce, yu2024blueprinting, kraska2023check, saxena2023auto}. However, spinning up new clusters can require non-negligible time, so purely reactive approaches will transiently leave the system under-provisioned, risking SLO violations.

Finally, existing systems route queries using static workload classes, query templates, or estimated resource requirements~\cite{ortiz2018slaorchestrator, yu2024blueprinting, marcus2016wisedb, saxena2023auto, nathan2024intelligent}. They do not  \textbf{react} to live concurrent query load by explicitly considering how placing a new query on a particular cluster affects SLO adherence (for both the new query and the already-running ones). This is how RAIS~\cite{nathan2024intelligent} still falls short of addressing the full problem: since it only exposes a \emph{qualitative} price-performance slider, rather than a way to specify latency SLOs, it cannot provide SLO-aware query placement.

\sparagraph{Our Approach} \thesystem addresses the gaps discussed above by jointly providing all three key desired behaviors: planning from historical workload patterns, adjusting resources when reality deviates from the plan, and reacting to live concurrent load when routing each query. It does so through three corresponding components: the \component{Policy Tuner}, the \component{Autoscaler}, and the \component{Query Router}.

The \component{Policy Tuner} can \textbf{plan} using the historical workload.  It generates and simulates forecasted workloads to determine proactive cluster-management actions and configure parameters. By ensuring that the right resources are available at the right time (shortly before demand materializes), it reduces dependence on reactive scaling and mitigates cluster spinup delays.

The \component{Autoscaler} can \textbf{adjust} the active cluster set when observed workload behavior deviates from the forecast. Its important contributions include cluster spinup/teardown triggers, which identify opportune moments for resource adjustments, and a cluster spinup size selector, which uses short-term simulations to estimate the SLO and cost implications of different scaling actions.

Finally, the \component{Query Router} can \textbf{react} to live load. When a query $q$ arrives, it selects on which active cluster it should be executed by managing a multi-way tradeoff among \begin{enumerate*}
    \item the risk that $q$ will violate its SLO;
    \item  the risk $q$ poses to the SLO adherence of already-running queries; and
    \item the resource cost implied by each routing decision. 
\end{enumerate*}

\sparagraph{Contributions} In summary, we make the following contributions:

\begin{itemize}[leftmargin=*]

\item We formulate the problem of SLO-aware multi-cluster workload management for cloud data warehouses, where the goal is bounded SLO violation rate at minimal infrastructure cost.

\item We present \thesystem, a multi-timescale framework that provides the three key desired behaviors needed for this problem: planning from historical workload patterns, adjusting resources when workloads deviate from the forecast, and reacting to live concurrent load when routing individual queries.

\item We develop the components of \thesystem: a forecast-driven \component{Policy Tuner} for proactive cluster management and  tuning, a simulation-based \component{Autoscaler} for cost- and SLO-aware scaling, and a concurrency-aware \component{Query Router} for per-query placement.

\item We evaluate \thesystem on realistic Redbench-generated workloads and show that it can successfully meet latency SLOs of varying strictness, reducing cost by a mean of $26.4\%$ compared to the per-scenario next-best baseline. 

\item We also evaluate per-component performance and find that the \component{Query Router} and \component{Autoscaler} reduce the SLO violation rate by a mean of $47.8\%$ and $93.7\%$ respectively, compared to corresponding alternatives on TPC-DS-derived workloads, while the \component{Policy Tuner} can reduce the SLO violation rate by a mean of $44.6\%$ using a single day of workload history. Each component is also shown to be efficient enough for its operating timescale.

\end{itemize}

\section{Problem Formulation}~\label{sec:defs}

We now define the workload and system setting more precisely.

\sparagraph{Online, Diverse Query Arrivals} Each query $q$ is issued online at time $q.t_a$. At each point in time, no information about future query arrivals is known exactly. Each query $q$ has varying resource demands (e.g. CPU, memory, I/O), which may interact with those of concurrently executing queries, impacting their latency.

\sparagraph{Latency SLOs} Each query $q$ has a latency service-level objective $SLO(q)$, treated as an input provided by the application or administrator. For example, the SLO may be inferred from the query's source application, user-defined priority, or other submission-time metadata. A query $q$ has \emph{met} it whenever its \emph{client-side} latency $L(q) \leq SLO(q)$. Otherwise, $q$ has \emph{violated} its SLO.

\sparagraph{Clusters} Multiple clusters can query the same data, with the vendor managing concurrency control and consistency. Each cluster $c$ has a \emph{size} $S(c)$ describing its computational capacity. The queries running on $c$ are denoted $Q_c$. Instead of statically assigning each workload to a cluster, we allow routing queries independently:

\begin{defbox}{def:routing-policy}{Routing Policy}
    Let $\mathcal{C}_{q.t_a}$ be the active cluster set when query $q$ arrives, and $\mathcal{H}_{q.t_a}$ be the historical workload observed until then, including prior query arrivals, routing decisions, completions, and SLOs. A routing policy $\mathcal{R}$ determines which active cluster to execute $q$ on:
    $$\mathcal{R}(q, \mathcal{C}_{q.t_a}, \mathcal{H}_{q.t_a}) \mapsto c \in \mathcal{C}_{q.t_a}$$
\end{defbox}

\sparagraph{Adjustable Cluster Pool} New clusters can be requested from the vendor, and existing clusters can be released. Spinning up a new cluster $c$ takes time $\delta$; tearing down an existing cluster takes time $\zeta$. We present $\delta$ and $\zeta$ as constants, but our techniques also support sampling them. This leads to another decision point:

\begin{defbox}{def}{Scaling Policy}
At time $t$, given the active cluster set $\mathcal{C}_t$ and the historical workload $\mathcal{H}_t$, a scaling policy $\mathcal{S}$ outputs a new active cluster set $\mathcal{C}_{new}$:
$$
\mathcal{S}(t, \mathcal{C}_t, \mathcal{H}_t) \mapsto \mathcal{C}_{new}
$$
\end{defbox}

\sparagraph{Usage-Based Billing} Let $\mathcal{P} = (\mathcal{R}, \mathcal{S})$ denote a pair of policies. The cost $k^\mathcal{P}(c, \mathcal{Q})$ of a cluster $c$ over workload $\mathcal{Q}$ with these policies is the product of: (i) A constant \emph{price} $p$; (ii) $c$'s size $S(c)$; and (iii) $c$'s \emph{active time} while $\mathcal{Q}$ is executed, measured by a timer as follows:

\begin{itemize}
    \item The timer is initially paused.
    \item If paused, it starts running when a query arrives at $c$.
    \item If running, it pauses when both are true: (i) there are no running queries on $c$; and (ii) at least $m$ seconds (often $m=60$~\cite{snowflake-pricing, aws-redshift-pricing}) have elapsed since the timer started running. 
\end{itemize}

Billing interacts with routing: using an idle cluster restarts its timer, while consolidating reduces cost but can cause interference.

\sparagraph{Target-Aware Comparison} For a workload $\mathcal{Q}$ under policies $\mathcal{P}$, let the \emph{SLO violation rate} as $ v^\mathcal{P}(\mathcal{Q}) = \frac{1}{|\mathcal{Q}|}\sum_{q \in \mathcal{Q}}\mathbf{1}[L^\mathcal{P}(q) > SLO(q)]$ and the \emph{total cost} as $k^\mathcal{P}(\mathcal{Q}) = \sum_c k^\mathcal{P}(c, \mathcal{Q}) $. We also allow specifying an SLO violation rate \emph{target} $\tau$. Not treating $\tau$ as a hard constraint allows comparing policies even when their SLO violation rate is above $\tau$. In particular, we compare policies with SLO violation rates $V = \{v^{\mathcal{P}_1},\ldots,v^{\mathcal{P}_n}\}$ and costs $K = \{k^{\mathcal{P}_1},\ldots,k^{\mathcal{P}_n}\}$ using:
$$ \textsc{BestWithTarget}(V,K,\tau) = \operatorname*{min}_{(v, k) \in \{V, K\}}^{\mathrm{lex}} \left( \max(0,\;v - \tau),\; k \right) $$

\sparagraph{Problem Definition} We want to find the best policy under the lexicographic target-aware objective above. Concretely: 

\begin{defbox}{def:problem}{SLO-aware Multi-Cluster Workload Management} 
 For an SLO violation rate target $\tau\geq 0$, select a policy pair $\mathcal{P}^\star$ to meet $\tau$ first and reduce cost second: 
 \[ \mathcal{P}^\star = \{\mathcal{P} \mid (v_\mathcal{P}, k_\mathcal{P}) = \textsc{BestWithTarget}(V,K,\tau) \} \] 

\end{defbox}

\begin{figure}[t]
    \centering
    \includegraphics[width=\linewidth]{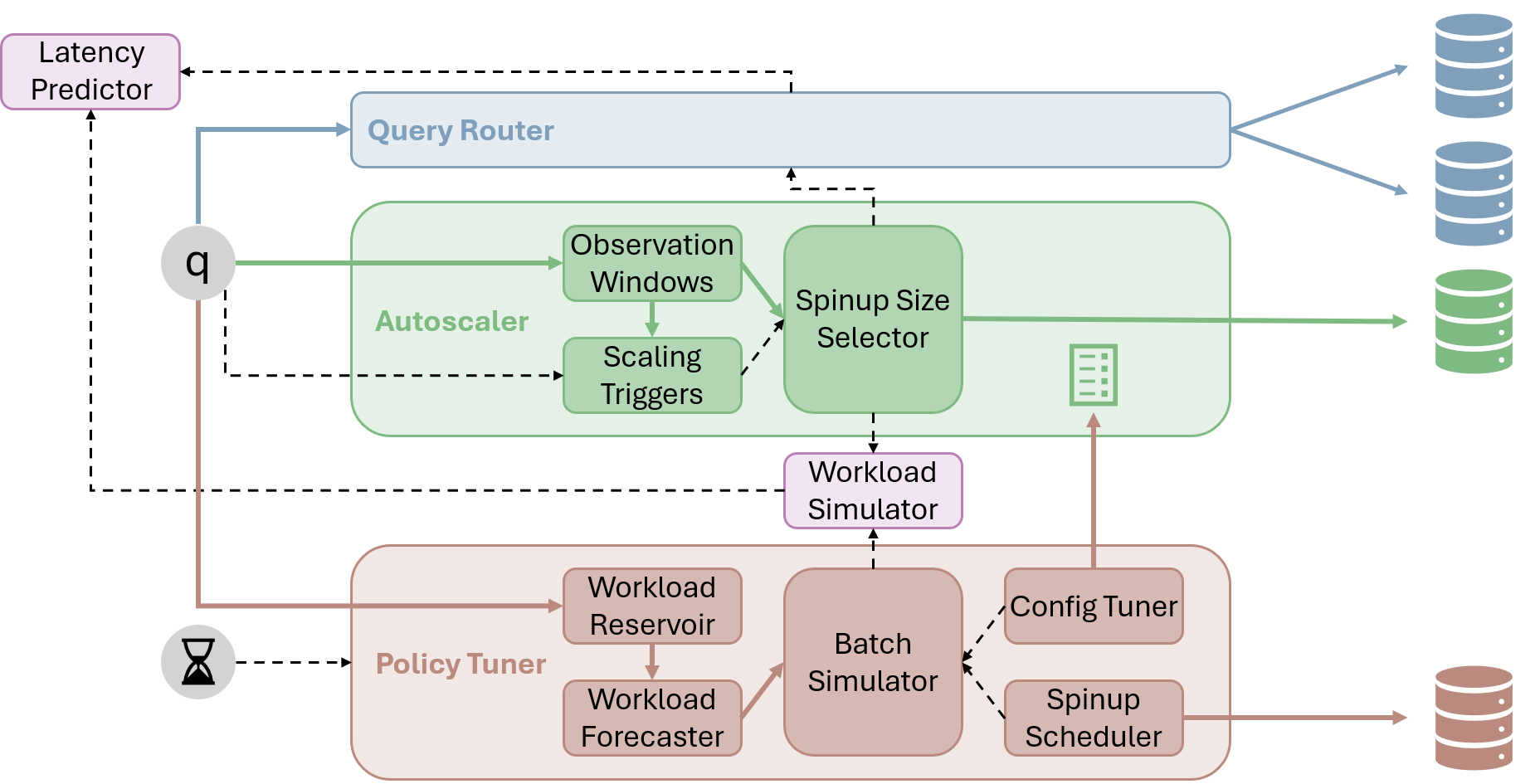}

    \caption{The architecture of \thesystem. Solid lines indicate input/output flow, while dashed lines indicate control flow.}
    \label{fig:architecture}
    \Description{An architectural diagram of \thesystem. The \component{Query Router} is a single blue box, invoking the \component{Latency Predictor}. The \component{Autoscaler} is a green box, with three sub-components: the \component{Observation Windows}, the \component{Scaling Triggers} and the \component{Spinup Size Selector}. Finally, the \component{Policy Tuner} is a red box with five sub-components: the \component{Workload Reservoir}, the \component{Workload Forecaster}, the \component{Batch Simulator}, the \component{Spinup Scheduler} and the \component{Configuration Tuner}. The \component{Spinup Size Selector} and the \component{Batch Simulator} both rely on the \component{Workload Simulator}, which invokes the \component{Latency Predictor}.}
\end{figure}

\section{Overview of \thesystem}\label{sec:overview}

To address the problem above, we design \thesystem to demonstrate the three \textbf{key desired behaviors} from Section~\ref{sec:introduction}. For our technical exposition in the following Sections, unlike our top-down introduction, we follow a query's path through the system bottom-up.

As shown in  Figure~\ref{fig:architecture}, an incoming query $q$ first encounters the \component{Query Router}, which can \textbf{react} to the current workload and decide which active cluster to forward $q$ to. This decision is powered by the \component{Latency Predictor}, which estimates how different placements would affect both $q$ and the already-running queries. Sections~\ref{sec:model} and~\ref{sec:query-router} describe the predictor and routing algorithm, respectively.

Each query is also added to the \component{Autoscaler}'s sliding \component{Observation Windows} and trips the \component{Scaling Triggers}, which decide whether to \textbf{adjust} the active cluster set. If a spinup is triggered, the \component{Spinup Size Selector} simulates a short-term workload forecast under different hypothetical cluster additions to balance SLO adherence and cost. It uses the \component{Workload Simulator}, a simple event-driven simulator that can replay query arrivals and uses the \component{Latency Predictor} to determine query completions. Section~\ref{sec:autoscaler} expands on this process.

Finally, each query is added to the \component{Workload Reservoir} within the \component{Policy Tuner}, which is triggered periodically (e.g. every 24 hours). The \component{Workload Forecaster} uses the reservoir to generate forecasted workloads, which are efficiently simulated using the \component{Batch Simulator}. Based on these simulations, the  \component{Spinup Scheduler} optimizes cluster spinup timing for reliably recurring load, while the \component{Configuration Tuner} then tunes \component{Autoscaler} parameters governing on-demand autoscaling.  Section~\ref{sec:policy-tuner} covers this functionality.

\section{Latency Predictor}\label{sec:model}

We start with Iconq~\cite{wu2025improving}, a recent LSTM-based concurrent query latency estimation model, but modify it to address the needs of \thesystem. In particular, we need support for clusters of different sizes (in the \component{Query Router}), while we also plan to use the model to determine query completion times during simulations (in the \component{Workload Simulator}). Before discuss how we address these requirements in our \component{Latency Predictor}, called Iconq+, we present Iconq.

\subsection{Background: Iconq}\label{sec:model-background}

Iconq was developed for single-cluster query scheduling, where arriving queries can be queued and/or reordered to reduce mean latency. At a high level, Iconq predicts the latency of a query $q$ by ingesting a sequence of \emph{interaction feature vectors}, derived from $q$ and its neighbors, with an LSTM. We define $q_n$ to be a neighbor of $q$ if their executions overlap. Each interaction feature vector concerns the target query $q$ and a particular neighbor $q_n$ and includes: query-plan-derived features for $q$ and $q_n$; relative timing information about their arrivals; and concurrency-unaware latency estimates for each query, $\hat{\ell}_q$ and $\hat{\ell}_{q_n}$, derived from a fast decision-tree-based sub-model (Stage~\cite{wu2024stage}). Importantly, Iconq ingests neighbors arriving before $q$ and after $q$ separately, enabling latency prediction updates for already-running queries to account for  submissions of new ones.

\begin{figure}
    \centering
    \includegraphics[width=\linewidth]{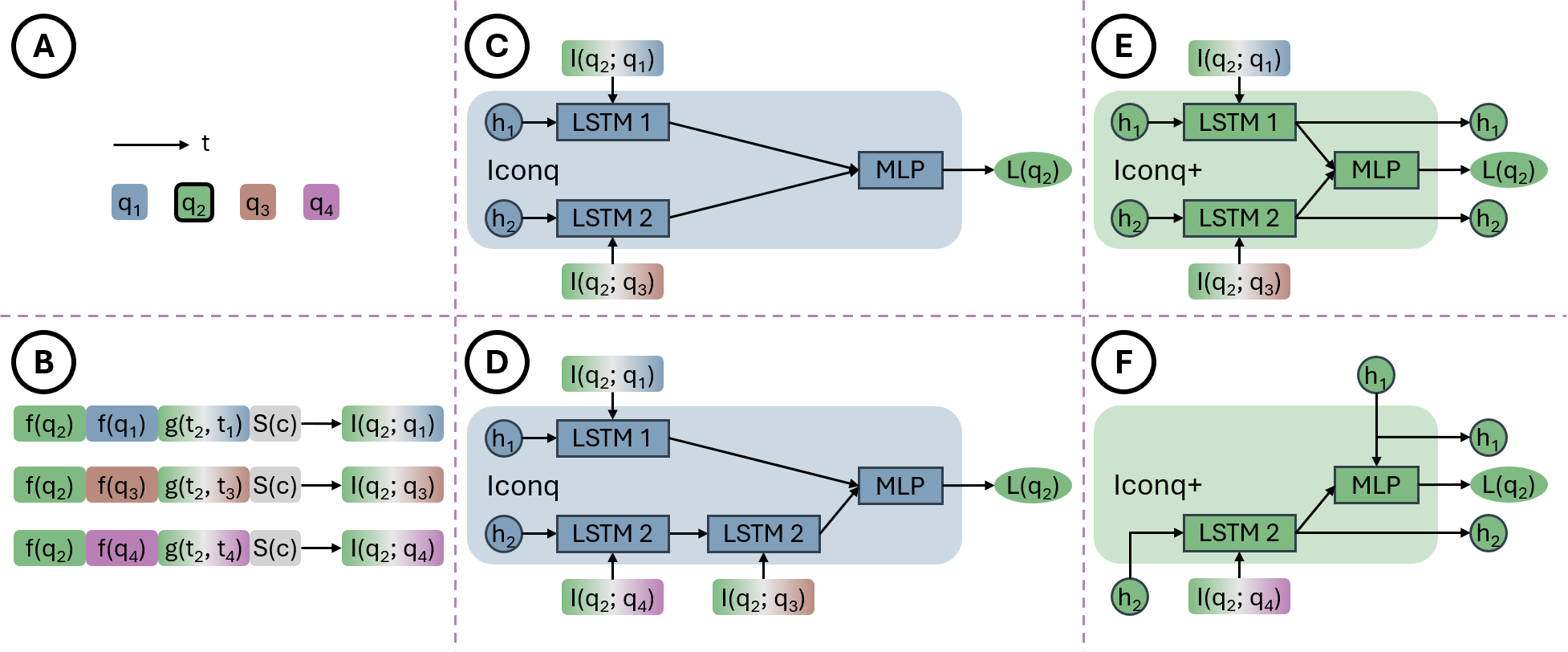}

    \caption{When predicting the latency of $q_2$, Iconq+ ingests the interaction feature vectors of later neighbors ($q_3$ and $q_4$) in chronological order, enabling incremental inference.}
    \label{fig:model}

    \Description{(A): Four queries, $q_1$ through $q_4$, arriving in chronological order, with $q_2$ highlighted. (B): Three interaction feature vectors, each of which covers a pair of queries. $q_2$ is always the first element of each pair. Each vector has individual-query features for each of its constituent queries, plus relative timing features and cluster size features. (C): Iconq after $q_1$--$q_3$ have arrived; to predict the latency of $q_2$, the interaction feature vector between $q_1$ and $q_2$ is ingested in the forward LSTM pass (because $q_1$ arrived before $q_2$), while the interaction feature vector between $q_3$ and $q_2$ is ingested in the backward LSTM pass. (D): When $q_4$ arrives, Iconq has to repeat the entire backward pass to update its prediction for $q_2$, because its architecture must ingest the interaction feature vector between $q_4$ and $q_2$ before the interaction feature vector between $q_3$ and $q_2$ in the backward pass. (E): Iconq+ after $q_1$--$q_3$ have arrived; like Iconq, to predict the latency of $q_2$, the interaction feature vector between $q_1$ and $q_2$ is ingested in the forward LSTM pass (because $q_1$ arrived before $q_2$), while the interaction feature vector between $q_3$ and $q_2$ is ingested in the backward LSTM pass. (F): When $q_4$ arrives, Iconq+ only has to ingest the interaction feature vector between $q_4$ and $q_2$, because it ingests the after-neighbor vectors in chronological order.}
\end{figure}

\subsection{Query Featurization}\label{sec:model-size}

Iconq assumes all queries run on the same cluster. In \thesystem, however, we must predict latencies across clusters of different sizes. We therefore extend the \emph{interaction feature vector} with features related to cluster size, including: (i) the raw cluster size; (ii) $\log_2(\mathrm{size})$ to capture non-linear scaling; (iii) $1/\mathrm{size}$ to capture inverse-capacity effects; (iv) $\hat{\ell}_q \cdot \mathrm{size}$ and $\hat{\ell}_{q_n} \cdot \mathrm{size}$ as proxies for the compute work implied by each query's isolated latency; and (v) $(\hat{\ell}_q + \hat{\ell}_{q_n})/\mathrm{size}$ as a capacity-normalized pairwise contention signal. 
In Figure~\ref{fig:model}\textbf{B}, we see the interaction feature vectors for the queries in Figure~\ref{fig:model}\textbf{A}, when predicting the latency of $q_2$. Each includes query plan features and Stage latency predictions per query $f(q)$, relative timing features $g(t;t_{neighbor})$ and cluster size features $S(c)$ as above.

\subsection{Censored Observations}\label{sec:model-censored}

To predict the latency of $q$, Iconq uses features derived from $q$ and its neighbors. When scheduling online, precise neighbor sets are known at decision time. In \thesystem, however, we also use the \component{Latency Predictor} in the \component{Workload Simulator}. There, whether two queries overlap can depend on their simulated (predicted) latencies, which require overlap knowledge. Hard inclusion of neighbors therefore creates a feedback loop, where early prediction errors alter downstream neighbor sets and lead to compounding errors.

To mitigate this, we also train Iconq+ on censored observations using sampled partial neighbor sets, where labels are treated as \emph{lower bounds} (i.e. loss is zero if $pred  \geq label$). This lets us extract more from the training data. For example, if $q$ ran from $t=0$ to $t=10$ and $q'$ arrived at $t=7$, then we know that $L(q)=10$ in the presence of $q'$; however, we also know that $L(q) \geq 7$  without $q'$, since $q$ was still running when $q'$ arrived. 

Censoring also lets us train on queries aborted due to timeouts or errors, with time-to-failure as their censored label. This is especially important for small clusters, where timeouts can be more common; discarding timed-out queries would over-represent successful executions and bias the model toward latency under-prediction. 

\subsection{Incremental Predictions}\label{sec:model-incremental}

Iconq uses a bidirectional LSTM to ingest a sequence of interaction feature vectors. Per Figure~\ref{fig:model}\textbf{C}/\textbf{D}, when predicting the latency of $q_2$, neighbors arriving before $q_2$ (i.e. $q_1$) are ingested in chronological order, while neighbors arriving after $q_2$ (i.e. $q_3$ and $q_4$) are ingested in reverse chronological order. This makes both sequences end near $q_2$'s arrival, emphasizing nearby interactions in the LSTM state. However, it also means that whenever a new neighbor arrives (e.g. $q_4$ in panel \textbf{D}), every neighbor after $q_2$ must be re-ingested.

Although the absolute re-ingestion overhead per query can be small, it can add up during simulations, increasing the runtime of the \component{Autoscaler} and the \component{Policy Tuner}. This is especially true because simulations interleave predictions and CPU-heavy state bookkeeping, effectively mandating CPU inference.  To avoid repeated re-ingestion, we modify Iconq+ to also process after-neighbors in chronological order, and add support for incremental inference from cached model state (Figure~\ref{fig:model}\textbf{E}/\textbf{F}).

\section{Query Router}\label{sec:query-router}

\begin{algorithm}[t]
\caption{Overall workflow of the \component{Query Router} (\textsc{Route}).}
\label{alg:query-router}
\begin{algorithmic}[1]
\item[\textbf{Inputs:}] Incoming query $q$, active cluster set $\mathcal{C} = \{c\}$ with  latency predictions $\hat{L}_c^{\text{before}}$ and latency predictor states $M_c^{\text{before}}$ for running queries.
\item[\textbf{Outputs:}] Selected cluster $c^\star$, updated latency predictions $\hat{L}_{c^\star}$ and latency predictor states $M_{c^\star}$
\For{$c \in \mathcal{C}$}
    \State $v_c^{\text{before}} \gets \sum_{q_i \in Q_c} \!\!\mathbf{1}[\hat{L}_c^{\text{before}}(q_i)>\mathrm{SLO}(q_i)]$
    \State $k_c^{\text{before}} \gets \Call{ClusterCostUntilDrained}{c, Q_c, \hat{L}_c^{\text{before}}}$
    \State $\mathcal{N}_c \gets \Call{PerQueryNeighborsIfWeRoute}{c, q}$
\EndFor

\State $\{L_c^\text{after}, M_c^\text{after}\}_{c \in \mathcal{C}} \gets \Call{PredictWithIconq+}{\{\mathcal{N}_c, M_c^\text{before}\}_{c \in \mathcal{C}}}$
\For{$c \in \mathcal{C}$}
    \For{$q_i \in Q_c$}
        \State $\hat{L}^\text{after}_c(q_i) \gets \max\!(\hat{L}^\text{after}_c(q_i),\, \hat{L}_c^{\text{before}}(q_i), q.t_\text{arrival}- q_i.t_\text{arrival})$
    \EndFor
\EndFor

\For{$c \in \mathcal{C}$}
    \State $Q_c^{\text{after}} \gets Q_c \cup \{q\}$
    \State $v_c^{\text{after}} \gets \sum_{q_i \in Q_c^{\text{after}}}\!\!\mathbf{1}[\hat{L}_c^{\text{after}}(q_i)>\mathrm{SLO}(q_i)]$
    \State $k_c^{\text{after}} \gets \Call{ClusterCostUntilDrained}{c, Q_c^{\text{after}}, \hat{L}^\text{after}_c}$
    \State $\Delta v_c \gets  v_c^{\text{after}} - v_c^{\text{before}}$
    \State $\Delta k_c \gets k_c^{\text{after}} - k_c^{\text{before}}$
\EndFor

\State $c^\star \gets \operatorname*{lexicographic-argmin}_{c\in\mathcal{C}}\; \bigl(\Delta v_c,\, \Delta k_c\bigr)$ 
\State \Return $c^\star$, $\hat{L}_{c^\star}^\text{after}$, $M_{c^\star}^\text{after}$

\end{algorithmic}
\end{algorithm}

\newcommand{\figheight}{3cm}

The \component{Query Router} routes each incoming query $q$ to an active cluster, leveraging the \component{Latency Predictor} to balance three concerns: (1) the ability of $q$ to meet its SLO, (2) the impact of $q$ on the SLO adherence of currently running queries, and (3) the cost implications of the routing decision. 
The \component{Query Router} implements \textsc{Route} (Algorithm~\ref{alg:query-router}). It first inspects each cluster $c$, gathering the SLO-violating queries (line 2) and the total projected cost, from $c$'s spinup until the running queries are predicted to drain (line 3). In the same pass, it builds each cluster's counterfactual query neighbor map, as if the incoming query $q$ were placed there (line 4).

It then runs Iconq+ inference on these counterfactual neighbor maps in a single batch (line 5), yielding updated model predictions $\hat{L}_c^\text{after}$  and model states $M_c^\text{after}$. As explained in Section~\ref{sec:model-incremental}, Iconq+ updates the latency predictions and model states for already-running queries \emph{incrementally} within this step. \textsc{Route} then applies a monotonicity guard for stability (lines 6-8): the latency prediction for each running query is never updated downwards, reflecting the fact that later queries generally cannot make an earlier query faster, and cannot be smaller than its observed latency so far.

For each candidate cluster $c$, the algorithm then counts the SLO-violating queries and implied cost in a manner similar to the baseline state (lines 9-12), ultimately calculating the marginal impact of the routing decision (lines 13-14).
It then identifies the cluster $c^\star$ that minimizes marginal SLO violations and cost (in this order) and returns it, together with the corresponding updated latency predictions and latency predictor states (lines 15-16).

\begin{figure}[t]

\centering

\begin{subfigure}{0.24\linewidth}
    \centering
    \includegraphics[width=\linewidth,height=\figheight, keepaspectratio]{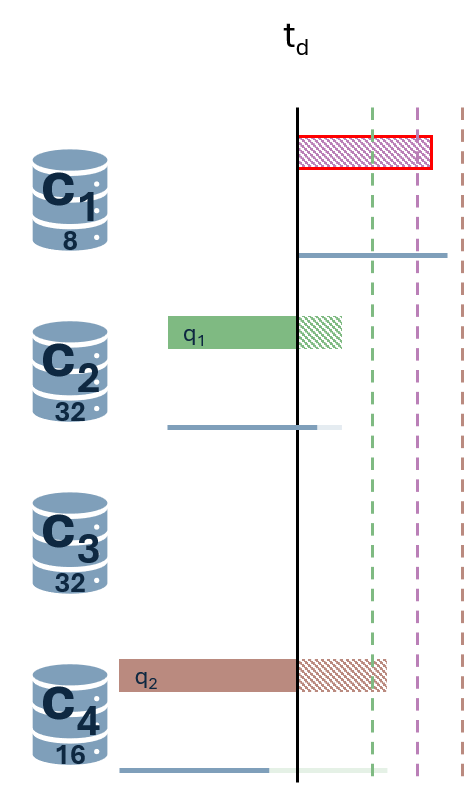}
    \caption{$q$ SLO miss}
    \label{fig:query-router-1}
\end{subfigure}
\hfill
\begin{subfigure}{0.24\linewidth}
    \centering
    \includegraphics[width=\linewidth,height=\figheight, keepaspectratio]{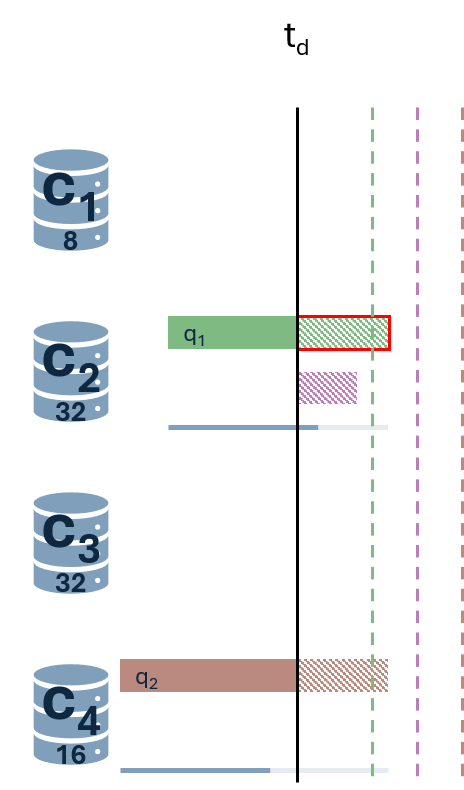}
    \caption{$q_1$ SLO miss}
    \label{fig:query-router-2}
\end{subfigure}
\hfill
\begin{subfigure}{0.24\linewidth}
    \centering
    \includegraphics[width=\linewidth,height=\figheight, keepaspectratio]{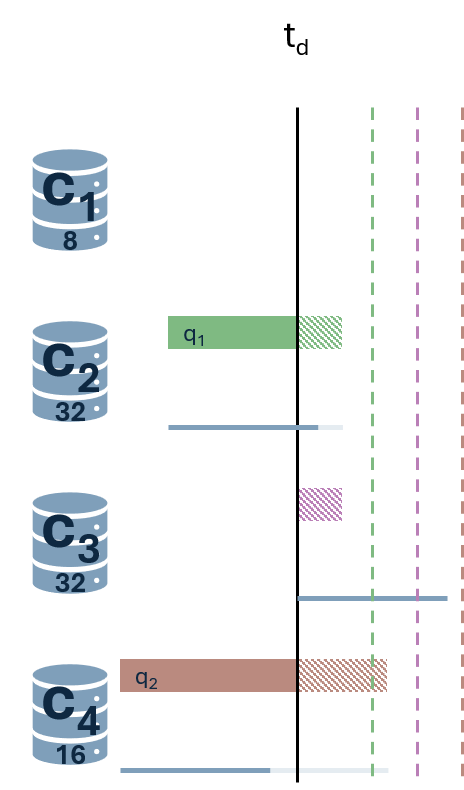}
    \caption{Expensive}
    \label{fig:query-router-3}
\end{subfigure}
\hfill
\begin{subfigure}{0.24\linewidth}
    \centering
    \includegraphics[width=\linewidth,height=\figheight, keepaspectratio]{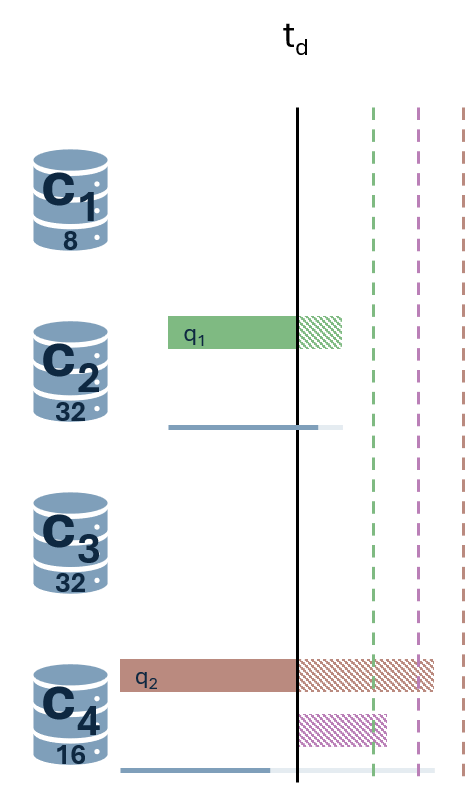}
    \caption{Balanced}
    \label{fig:query-router-4}
\end{subfigure}

\caption{The \component{Query Router} balances SLOs and cost. The x-axis is time. Each query is a rectangle and same-color dashed lines show when it will miss its SLO. Hasing indicates predicted latencies; a red outline marks an SLO miss. Thin blue lines show billed time; dark blue indicates the minimum.}

\Description{Please refer to the text of Section~\ref{sec:query-router} for a Figure walkthrough.}
\label{fig:query-router}

\end{figure}

Figure~\ref{fig:query-router} shows an example, where query $q$ (purple) is routed among 4 clusters, with 2 active queries $q_1$ (green) and $q2$ (brown). 
Routing $q$ to $c_1$ would make it miss its SLO, while routing it to $c_2$ would make $q_1$ miss its own SLO. Routing it to $c_3$ would imply no SLO violations, but it would start a new billing window, leading to high cost. The best option is therefore $c_4$; there are no SLO violations, while the marginal cost impact is minimized.

 By explicitly reasoning about query interference, SLO violation risk, and resource cost, the \component{Query Router} can make  decisions that balance performance and efficiency in real time.

\sparagraph{Algorithm~\ref{alg:query-router} Computational Time Complexity} Before and after model inference, \textsc{Route} performs a constant amount of work per running query, implying $O(N)$,  where $N = \sum_c  |Q_c|$. For line 5, each LSTM step is $O(dh + Lh^2)$ (to ingest the input and update the hidden state per layer), assuming input dimension $d$,  hidden size $h$ and $L$ layers. We will treat this as constant, since we use a fixed model. We perform two steps per running query $q_\text{running}$: one to update $q_\text{running}$'s latency prediction if $q$ is added to its neighbors (see Section~\ref{sec:model-incremental}), and another while ingesting $q_\text{running}$ as a neighbor of $q$ while predicting the latency of $q$. Line 5 is then also $O(N)$, meaning that \textsc{Route}'s overall time complexity is $O(N)$.

\begin{algorithm}[t]
\caption{The spinup trigger of the \component{Autoscaler} (\textsc{MaybeSpinup}).}
\label{alg:autoscaler-spinup-trigger}
\begin{algorithmic}[1]
\item[\textbf{Inputs:}] Completion window $\mathcal{W}^c$, active cluster set $\mathcal{C} = \{c\}$  with predictions $\hat{L}_c$
\item[\textbf{Configuration:}] Trigger SLO threshold $\tau_\text{trigger}$, minimum observations $\theta$
\item[\textbf{State:}] Known active cluster set $\mathcal{K}$, change bound $t_{\mathrm{change}}$ (shared with \textbf{Algorithm~\ref{alg:autoscaler-sim}}), in-flight flag $f$ (shared with \textbf{Algorithm~\ref{alg:autoscaler-sim}})
\item[\textbf{Outputs:}]  Whether to invoke the \component{Spinup Size Selector}.

\If{$\mathcal{K} \neq \mathcal{C}$}
\State  $\mathcal{K} \gets  \mathcal{C}$
\State $t_{\mathrm{change}} \gets \max(t_{\mathrm{change}}, q.t_{\mathrm{arrival}})$
\State $f \gets \textsc{False}$
\EndIf
\State $R \gets \{(q, \hat{L}_c(q)) \mid (q.t_{\mathrm{arrival}} \ge t_{\mathrm{change}}) \land (q\in Q_c,  c \in \mathcal{C})\}$
\State $D \gets \{(q, L(q)) \mid (q.t_{\mathrm{arrival}} \ge t_{\mathrm{change}}) \land (q\in \mathcal{W}^c)\}$
\If{$\lnot f$ \textbf{ and } $|R| + |D| \ge \theta$}
    \State $v_{\mathrm{cur}} \gets \frac{1}{|R| + |D|}\sum_{(q, \ell)\,\in\,\{R \cup D\}} \mathbf{1}[\ell>\mathrm{SLO}(q)]$
    \If{$v_{\mathrm{cur}} > \tau_\text{trigger}$}
        \State $f \gets \textsc{True}$
        \State \Return \textsc{True}
    \EndIf
\EndIf

\State \Return \textsc{False}
\end{algorithmic}
\end{algorithm}

\section{Autoscaler}\label{sec:autoscaler}

The \component{Query Router} routes each query \emph{among the active cluster set} $\mathcal{C}$; however, it may be that $\mathcal{C}$ is no longer appropriate. The \component{Autoscaler} addresses this, using three sub-components (see Figure~\ref{fig:architecture}).

\subsection{Observation Windows}\label{sec:autoscaler-windows}

The \component{Autoscaler} runs in a background thread and maintains two trailing \component{Observation Windows} over the last $w$ seconds. It appends each incoming query to the \emph{arrival window} $\mathcal{W}^a$ while it appends each completed query (alongside its arrival time and latency) to the \emph{completion window} $\mathcal{W}^c$, truncating each as needed. Each query arrival also trips two \component{Scaling Triggers}, explained next.

\subsection{Scaling Triggers}\label{sec:autoscaler-triggers}

\subsubsection{Teardown Trigger}\label{sec:autoscaler-triggers-teardown}

The teardown trigger determines whether one of the active clusters should be torn down. For an active cluster $c$, it fires when three conditions are simultaneously met: (i) no queries are currently running on $c$; (ii) no query has been routed to $c$ for the last $T_\text{idle}$ seconds; and (iii) $c$ was spun up at least $T_\text{min\_lifetime}$ seconds earlier. If all conditions are met, the \component{Autoscaler} initiates the teardown of $c$. $T_\text{idle}$ and $T_\text{min\_lifetime}$ can be periodically optimized by the \component{Policy Tuner}, as we will examine in Section~\ref{sec:policy-tuner-sweep}.

\subsubsection{Spinup Trigger}\label{sec:autoscaler-triggers-spinup}

The spinup trigger determines whether a new cluster should be spun up, based on whether recent SLO adherence is unsatisfactory. As described in Algorithm~\ref{alg:autoscaler-spinup-trigger}, it first checks whether the active cluster set has changed since the last invocation; if so, it saves it, records the current arrival time as $t_{\mathrm{change}}$ and clears the in-flight flag $f$ (lines~1--4). The algorithm then assembles $R$ (line~5), the set of running queries that arrived after $t_\text{change}$, and $D$ (line~6), the set of queries from $\mathcal{W}^c$ that arrived after $t_\text{change}$, alongside their predicted and realized latencies, respectively. If the combined size of these two sets is at least $\theta$ and the in-flight flag is not set (line 7), the algorithm computes the SLO violation rate among $R \cup D$ (line 8) and compares it to $\tau_\text{trigger}$ (line 9). The trigger fires only if this threshold is exceeded, at which point the in-flight flag is also set (lines 10-12). The parameters $\tau_\text{trigger}$ and $\theta$ can be periodically optimized by the \component{Policy Tuner}, as we will examine in Section~\ref{sec:policy-tuner-sweep}.
If the spinup trigger fires, the \component{Autoscaler} invokes the \component{Spinup Size Selector}, explained next in Section~\ref{sec:autoscaler-size-selector}.

\sparagraph{Algorithm~\ref{alg:autoscaler-spinup-trigger} Computational Time Complexity} The comparison on line 1 takes time $O(\min(|\mathcal{K}|, |\mathcal{C}|))$, while constructing $R$ and $D$ on lines 5-6 requires time $O(N + |\mathcal{W}_c|)$,  where $N = \sum_c  |Q_c|$. Line 8 (if reached) operates on a subset of these query sets, so the overall complexity is $O(\min(|\mathcal{K}|, |\mathcal{C}|)+ N + |\mathcal{W}_c|)$.

\subsection{Spinup Size Selector}\label{sec:autoscaler-size-selector}

\begin{algorithm}[t]
\caption{The \component{Spinup Size Selector} (\textsc{FindBestSpinupSize}).}
\label{alg:autoscaler-sim}
\begin{algorithmic}[1]
\item[\textbf{Inputs:}] Arrival window $\mathcal{W}^a$, active cluster set $\mathcal{C} = \{c\}$  with  latency predictions $\hat{L}_c$, decision time $t_d$, target SLO violation rate $\tau_\text{target}$       
\item[\textbf{Configuration:}] Eligible cluster sizes $\mathcal{S}$, spinup delay $\delta$, minimum completions $\mu$, observation window width $w$
\item[\textbf{State:}] Change bound $t_{\mathrm{change}}$, in-flight flag $f$ (both shared with \textbf{Algorithm~\ref{alg:autoscaler-spinup-trigger}})
\item[\textbf{Outputs:}]  Cluster size $s$ to spin up, or $\emptyset$ if none advisable.

\State $t_{\mathrm{available}} \gets t_d + \delta$;\quad $i \gets 0$
\While{\textbf{true}}
    \State $q \gets \mathcal{W}^a[i \mod |\mathcal{W}^a|]$
    \State $q.t_{\mathrm{arrival}} \gets q.t_{\mathrm{arrival}} + w\cdot \lfloor i / |\mathcal{W}^a| \rfloor$
    \If{$q.t_{\mathrm{arrival}} \ge t_{\mathrm{available}}$}
       
        \State \textbf{break} 
    \EndIf
    \State $\Call{CollectFinishedUntil}{\mathcal{C},\,q.t_{\mathrm{arrival}}}$ 
    \State $(c^\star,\,\hat{L}_{c^\star}, M_{c^\star}) \gets \Call{Route}{q,\;\mathcal{C}}$
    \State $c^\star.\Call{AddQuery}{q,\,\hat{L}_{c^\star}, M_{c^\star}}$
    \State $i \gets i+1$
\EndWhile
\State $(\mathcal{C}_\text{snap}, i_\text{snap}) \gets(\mathcal{C}, i)$
\State  $\mathcal{S}' \gets\mathcal{S} \cup \{\emptyset\}$
\For{$s \in \mathcal{S}'$}
    \State $\mathcal{C}_s \gets \mathrm{copy}(\mathcal{C}_{\mathrm{snap}})$
    \If{$s \neq \emptyset$}
        \State $\mathcal{C}_s \gets \mathcal{C}_{s} \cup \{\Call{NewCluster}{\mathrm{size}=s}\}$         
    \EndIf
    \State $i \gets i_\text{snap}$
    \State $D \gets \emptyset$
    \While{$|D| < \mu$}
       \State $q \gets \mathcal{W}^a[i \mod |\mathcal{W}^a|]$
        \State $q.t_{\mathrm{arrival}} \gets q.t_{\mathrm{arrival}} + w\cdot \lfloor i / |\mathcal{W}^a| \rfloor$
        \State $F \gets \Call{CollectFinishedUntil}{\mathcal{C}_s,\,q.t_{\mathrm{arrival}}}$
        \State $D \gets D \cup\{(q', L(q')) \mid (q'  \in F) \land  (q'.t_\text{arrival} \geq t_\text{available})  \}$
        \State $(c^\star,\,\hat{L}_{c^\star},\,M_{c^\star}) \gets \Call{Route}{q, \mathcal{C}_s}$
        \State $c^\star.\Call{AddQuery}{q,\,\hat{L}_{c^\star},\,M_{c^\star}}$
        \State $i \gets i + 1$
    \EndWhile
    \State $D \gets D \cup_{c\in \mathcal{C}_s} \{(q', \hat{L}_c(q')) \mid q' \in Q_c \}$
    \State $v_s \gets \frac{1}{|D|}\sum_{(q, \ell)\,\in\,D} \mathbf{1}[\ell>\mathrm{SLO}(q)]$
    \State $k_s \gets \sum_{c\in\mathcal{C}_s} \Call{ClusterCostUntilDrained}{c,\,Q_c,\,\hat{L}_c}$
\EndFor
\State $(v_\text{best}, k_\text{best}) \gets \Call{BestWithTarget}{\{(v_s, k_s)\}_{s \in \mathcal{S}'}, \tau_\text{target}}$ 
\If{$(v_\text{best}, k_\text{best}) = (v_\emptyset, k_\emptyset)$}
    \State $t_{\mathrm{change}} \gets t_d + \delta$
    \State $f \gets \textsc{False}$
    \State \Return $\emptyset$
\EndIf
\State \Return $\max\{s \in \mathcal{S} \mid (v_\text{best}, k_\text{best}) = (v_s, k_s)\}$ \

\end{algorithmic}
\end{algorithm}

Once the spinup trigger fires, the \component{Spinup Size Selector} must decide what cluster it would be most beneficial to spin up, if any. At a high level, for each eligible cluster size, it hypothesizes a cluster spinup of that size and compares the downstream SLO violation rate and cost. It does this by simulating a short-term workload forecast, where queries arrive according to copies of the arrival window $\mathcal{W}^a$, are routed as usual by the \component{Query Router}, and complete based on their predicted latencies. Crucially, it also evaluates a \emph{do-nothing baseline} (no new cluster added), and only recommends a spinup if some candidate strictly improves upon this baseline.

Remember that a new cluster will not be available instantly, but only after a spinup delay $\delta$. It is vital to capture this \emph{preparation period} in the simulation, because it may generate a workload backlog that the new cluster will have to help relieve, once available. However, the events of this preparation period are independent of the size of the requested cluster, so that it is sufficient to compute the simulation state at the end of the preparation period once.

More precisely, the \component{Spinup Size Selector} operates according to Algorithm~\ref{alg:autoscaler-sim}. Lines~1--11 describe the preparation period: copies of the queries in the arrival window are issued (lines~3--4 and 10) and routed (lines~8--9), while queries that are predicted to have completed are removed from tracking (line 7). Once the new cluster is available (lines ~1--2 and 5--6), replay stops and we snapshot the cluster set state (line 11). We extend the candidate set with the do-nothing baseline $\emptyset$ (line 12) and, for each possible action, initialize the cluster set from the snapshot (lines 13--17).

We then continue until we have simulated $\mu$ query completions (lines 18--19). The core simulation loop is mostly the same (lines 20--26), with the exception that we record the latency of completing queries that arrived after $t_\text{available}$ (lines 22--23). After exiting the simulation, we augment this set of $\mu$ completed queries with the predicted latencies of outstanding queries (line 27) and compute the SLO violation rate and cost (lines 28--29). We then find the action that provides the best balance of SLO violation rate and cost according to an SLO violation rate target $\tau_\text{target}$ (line 30, see the end of Section~\ref{sec:defs}).  If $\emptyset$ wins, we return it, suppressing trigger re-evaluation for $\delta$ seconds to prevent immediate re-firing (lines 31--34, see Algorithm~\ref{alg:autoscaler-spinup-trigger}). Otherwise, we return the largest cluster tied for the best outcome (line 35), erring on the side of caution when multiple sizes perform equally well. The parameters $\mu$ and $w$ can be periodically optimized by the \component{Policy Tuner}, as we will examine in Section~\ref{sec:policy-tuner-sweep}.

\begin{figure}
    \centering
    \includegraphics[width=\linewidth]{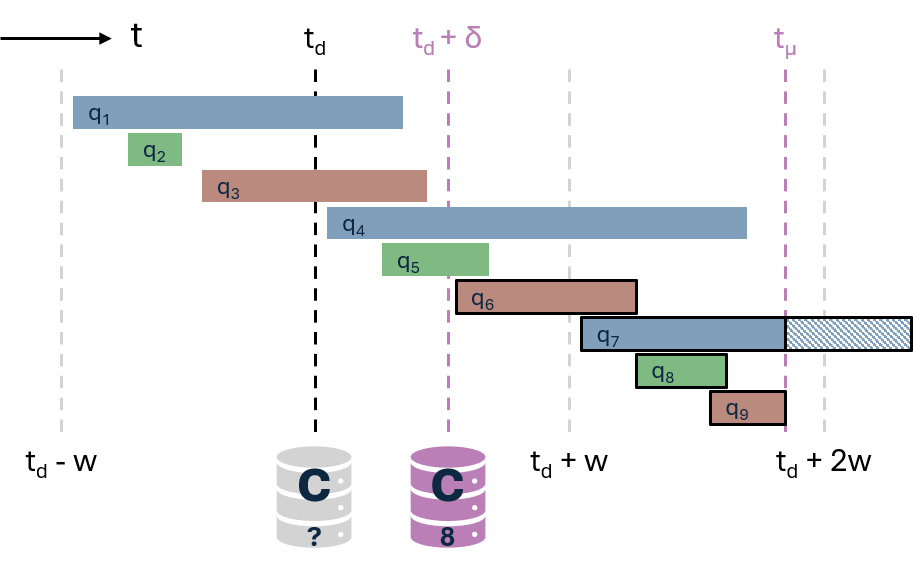}

    \caption{The \component{Spinup Size Selector} evaluates each scaling action at $t_d$ by simulating copies of the arrival window $\mathcal{W}^a$. It stops once  $\mu$  queries (here $\mu=3$) issued after the new cluster is available ($t_d + \delta$; queries outlined in black) finish executing.}
    \label{fig:autoscaler}

    \Description{Please refer to the text of Section~\ref{sec:autoscaler-size-selector} for a Figure walkthrough.}
\end{figure}

Figure~\ref{fig:autoscaler} shows an example for $\mu=3$, where the x-axis is time and each rectangle is a query. The query arrivals $q_1$--$q_3$ in the arrival window $\mathcal{W}^a$ are replayed twice until $q_9$'s completion at $t_{\mu}$ marks the third completion after $t_d + \delta$, the time the new (purple) cluster of size 8 became available. The simulated latencies of $q_6$, $q_8$ and $q_9$, and the predicted latency of $q_7$ as of $t_{\mu}$, contribute to $v_8$ and $k_8$.

\sparagraph{Algorithm~\ref{alg:autoscaler-sim} Computational Time Complexity} In the preparation phase (lines 2-10), the number of query arrivals will be $O(|\mathcal{W}^a|\cdot\frac{\delta}{w})$. Let the number of query arrivals required on lines 20-26 until $\mu$ queries complete be $F_s$ when the new cluster has size $s$ (or $s=\emptyset$), and let $F^\text{max} = \max_{s\in\mathcal{S}'}{F_s}$. Let each arriving query $q$ encounter $R_q$ running queries, among which any finished queries must be identified ($O(R_q)$) before $q$ is routed ($O(R_q)$), and let $R^\text{max} = \max_{q}{R_q}$. For the drain phase and SLO/cost calculations (lines 27-29) we similarly incur $O(R_q)$. Therefore, the overall complexity  of \textsc{FindBestSpinupSize} is $O((|\mathcal{W}^a|\cdot\frac{\delta}{w} + F^\text{max})\cdot R^\text{max})$.

\section{Policy Tuner}\label{sec:policy-tuner}

As described in Section~\ref{sec:autoscaler}, the \component{Autoscaler} can use the recent workload to adjust the active cluster. However, over longer time horizons, additional optimization opportunities open up. The \component{Policy Tuner}, invoked at the granularity of several hours\footnote{ In our exposition and experiments, we will assume that the \component{Policy Tuner} is invoked at the end of each day and optimizes over the day ahead, but this is not strictly required.}, captures such opportunities, as follows. Each incoming query is added to the \component{Workload Reservoir} (Section~\ref{sec:policy-tuner-reservoir}), which can be used by \component{Workload Forecaster} (Section~\ref{sec:policy-tuner-forecasting}) to draw forecasted workloads. The \component{Batch Simulator} (Section~\ref{sec:policy-tuner-batch-simulator}) can efficiently simulate these workloads under different configurations, empowering the \component{Spinup Scheduler} (Section~\ref{sec:policy-tuner-proactive}) to schedule proactive cluster spinups and the \component{Configuration Tuner} (Section~\ref{sec:policy-tuner-sweep}) to optimize \component{Autoscaler} knobs.

\subsection{Workload Reservoir}\label{sec:policy-tuner-reservoir}

Each query that reaches \thesystem is also added to the  \component{Workload Reservoir}. It consists of two tables indexed by calendar date and hour of day. The \emph{count table} records, for each $(date$, $hour$, $query\_template)$ triple, how many times that query template was issued. The \emph{arrivals table} records the within-hour second offset of every query arrival in each $(date,\,hour)$ bin. Both tables have configurable sampling policies to keep their storage footprint bounded, but they importantly do not need to be kept in memory, since they are only used by the infrequently invoked \component{Policy Tuner}.

\subsection{Workload Forecaster}\label{sec:policy-tuner-forecasting}

The \component{Workload Forecaster} generates forecasted workloads for the day ahead and partitions them into disjoint training and validation splits. In our current default implementation, it iterates over hours of the day and samples query texts from all available same-day-and-hour-of-week bins. It applies a decay to their sampling probability, so that samples from $k$ weeks ago are weighted by $\lambda^{k-1}$ (with $\lambda\in(0,1)$). To assign arrival times to each forecasted query, it calculates interarrival deciles from the \emph{arrivals table} (over the same-day-and-hour-of-week bins) and samples uniformly within each decile. For sparse bins, it falls back to uniform sampling. 

Note that the forecasted workloads only need to approximate the overall load contours of the real workload, so that coarse scaling decisions can be planned ahead of time. Alternative forecasting and sampling policies are also supported (e.g., a single prior day, or an equally-weighted 7-day window), but the described policy best balanced simplicity and predictive performance in our experience.

\subsection{Batch Simulator}\label{sec:policy-tuner-batch-simulator} 

\textsc{SimulateBatch} is the shared evaluation primitive invoked by every phase of the \component{Policy Tuner}. It accepts a set of \emph{execution configurations} $\{E_i\}$ and a set of workloads $W$, and returns a result matrix where entry $\mathcal{R}[i][j]$ holds the simulation outcome for configuration $E_i$ on workload $W[j]$. All $|\{E_i\}| \times |W|$ simulations are dispatched to a process pool and run in parallel in an optimized fashion. Each \emph{execution configuration} specifies any scheduled cluster spinups for the workload ahead (cluster size and spinup time), as well as values for the \component{Autoscaler} configuration parameters:  $T_\text{idle}$ and $T_\text{min\_lifetime}$ (Section~\ref{sec:autoscaler-triggers-teardown}); $\tau_\text{trigger}$ and $\theta$ (Section~\ref{sec:autoscaler-triggers-spinup}); and  $\mu$ and $w$ (Section~\ref{sec:autoscaler-size-selector}).

\subsection{Spinup Scheduler}\label{sec:policy-tuner-proactive}

\begin{algorithm}[t]
\caption{The \component{Spinup Scheduler} (\textsc{ScheduleSpinups}).}
\label{alg:spinup-optimizer}
\begin{algorithmic}[1]
\item[\textbf{Inputs:}]  Initial execution configurations $\mathcal{E} = \{E^0_i\}_{i\in \mathcal{I}}$, training workloads $W^\text{tr}$, validation workloads $W^\text{val}$, target SLO violation rate $\tau$ 

\item[\textbf{Configuration:}] Maximum spinups $\eta$, aggregation function $\alpha$, eligible cluster sizes $\mathcal{S}$, maximum attempts per round $\phi$

\item[\textbf{Outputs:}]  Optimized config $E^\star$ with scheduled spinups

\State $E^\text{best}_i \gets E^0_i$ for all $i \in \mathcal{I}$
\State $\text{alive\_at\_all} \gets \mathcal{I}$
\For{$r \in [0,\dots,\eta)$}
    \State $\{\Lambda_i\}_{i\in\text{alive\_at\_all}} \gets \Call{SimulateBatch}{W^\text{tr},\;\{E^\text{best}_i\}_{i\in\text{alive\_at\_all}}}$
    \For{$i \in \text{alive\_at\_all}$}
    \State $v_i^\text{base}, k_i^\text{base} \gets \Call{AggPerf}{\Lambda_i, \alpha}$ 
    \State $\Gamma_i \gets \Call{FindGoodSpinupTime}{\Lambda_i,\,\tau_\text{target}}$
    \If{$\Gamma_i = \emptyset$}
        $\text{alive\_at\_all} \gets \text{alive\_at\_all} \setminus \{i\}$
    \EndIf
    \EndFor
   
    \State $\text{attempt}_i \gets 0$ for all $i \in \text{alive\_at\_all}$
    \State $\text{alive\_this\_round} \gets \text{alive\_at\_all}$
    \While{$\text{alive\_this\_round} \neq \emptyset$}
        \For{$(i,\,s) \in \text{alive\_this\_round} \times \mathcal{S}$}
            \State $E_{i, s} \gets \Call{AddSpinup}{E^\text{best}_i,\;\text{time}=\Gamma_i(\text{attempt}_i),\,\text{size}=s}$
        \EndFor
        \State $\{\Lambda_{i, s}\} \gets \Call{SimulateBatch}{W^\text{tr},\;\{E_{i, s}\}}$
        \For{$i \in \text{alive\_this\_round}$}
            \State $v_{i, s}, k_{i, s} \gets \Call{AggPerf}{\Lambda_{i, s}, \alpha}$ for all $s \in \mathcal{S}$
            
            \State $E^\text{b}_i \gets \Call{BestWithTarget}{\{(v_{i, s},k_{i, s})\}_s \cup \{(v_i^\text{base},k_i^\text{base})\},\;\tau_\text{target}}$
            \If{$E^\text{b}_i \neq E^\text{best}_i$}
                \State $E^\text{best}_i \gets E^\text{b}_i$
                \State $\text{alive\_this\_round} \gets \text{alive\_this\_round} \setminus \{i\}$
            \ElsIf{$\text{attempt}_i + 1 < \min(|\Gamma_i|,\,\phi)$}
                \State $\text{attempt}_i \gets \text{attempt}_i + 1$
            \Else
                \State $\text{alive\_this\_round} \gets \text{alive\_this\_round} \setminus \{i\}$
                \State $\text{alive\_at\_all} \gets \text{alive\_at\_all} \setminus \{i\}$
            \EndIf
        \EndFor
    \EndWhile
\EndFor

\State $\{\Lambda_i^\text{val}\} \gets \Call{SimulateBatch}{W^\text{val},\;\{E^\text{best}_i\}_{i \in \mathcal{I}}}$
\State $v_i^\text{val}, k_i^\text{val} \gets \Call{AggPerf}{\Lambda_i^\text{val},\,\alpha}$ for all $i \in \mathcal{I}$
\State $E^\star \gets \Call{BestWithTarget}{\{(v_i^\text{val},\,k_i^\text{val})\}_{i\in \mathcal{I}},\;\tau}$
\State \Return $E^\star$

\end{algorithmic}
\end{algorithm}

\subsubsection{High-level Workflow}\label{sec:policy-tuner-proactive-high}

The \component{Spinup Scheduler} augments execution configurations with scheduled cluster spinups for periods of predictable congestion. Conceptually, it greedily evolves an initial execution configuration by adding spinups over at most $\eta$ rounds. In each round it simulates the current configuration on the training workloads, identifies promising times at which additional capacity may help reduce SLO violations, and accepts the best spinup candidate only if it improves upon the existing configuration.

Algorithm~\ref{alg:spinup-optimizer} implements this search for a set of initial configurations $\mathcal{E}=\{E_i^0\}_{i\in\mathcal{I}}$. It maintains the best configuration found so far for each candidate, $E_i^\text{best}$, and an index set of candidates that remain eligible for further spinups, $\text{alive\_at\_all}$ (lines~1--2). The outer loop performs up to $\eta$ greedy rounds (line~3). At the beginning of each round, all still-active candidates are simulated on the training workloads in a single batch (line~4). For each candidate, the scheduler aggregates its current SLO violation rate and cost (across the training workloads), then calls \textsc{FindGoodSpinupTime} (Algorithm~\ref{alg:find-spinup-time}) to obtain a ranked list $\Gamma_i$ of promising spinup times (lines~5--7). Candidates with no promising times are removed from $\text{alive\_at\_all}$ (line~8).

For the remaining candidates, the scheduler searches for one spinup to add in the current round. It initializes each candidate's attempt counter and places all active candidates into $\text{alive\_this\_round}$ (lines~9--10). The inner loop then evaluates candidates in attempt waves (lines~11--25). In each wave, every still-alive candidate $i$ is paired with every eligible cluster size $s\in\mathcal{S}$, producing a trial configuration that adds a spinup at the candidate's currently attempted time $\Gamma_i(\text{attempt}_i)$ with size $s$ (lines~12--13). These trial configurations are simulated in parallel on the training workloads, before we aggregate their violation rates and costs (lines~14--16).

The scheduler then produces, for each candidate, a local best $E_i^b$ among all new configurations and the current baseline using \textsc{BestWithTarget} (line~17). If it differs from the previous best, it is promoted to $E_i^\text{best}$ and the candidate leaves the current round because it has accepted a spinup (lines~18-20). If no cluster size improves the candidate at the current time, the scheduler advances to the next promising time in $\Gamma_i$, up to the per-round attempt limit $\phi$ (lines~21--22). If the candidate exhausts all allowed attempts without finding an improvement, it is removed both from the current round and from future rounds (lines~24--25). Thus, each outer round adds at most one spinup per candidate, and candidates stop being considered once no useful spinup can be found.

After the greedy training phase, we evaluate the best configuration for each initial candidate on the validation workloads (lines~26--27). We then select the configuration with the best validation cost subject to the target SLO violation rate $\tau$ and return it as $E^\star$ (lines~28--29). The batching in Algorithm~\ref{alg:spinup-optimizer} does not change the greedy search logic; it simply improves simulation throughput.

\sparagraph{Algorithm~\ref{alg:spinup-optimizer} Computational Time Complexity}
Let  $B$ the maximum parallel simulation batch size and $s$ the maximum simulation duration. Then the overall computational time complexity of \textsc{ScheduleSpinups} is
$O\!\left(|\mathcal{I}| \cdot \frac{s}{B} \cdot \left(\eta \cdot    |\mathcal{S}| \cdot \phi\cdot |W^\text{tr}| +  |W^\text{val}|\right)\right)$.

\subsubsection{Finding Good Scheduled Spinup Times}\label{sec:policy-tuner-time}

\begin{algorithm}[t]
\caption{Congestion point detection (\textsc{FindGoodSpinupTime}).}
\label{alg:find-spinup-time}
\begin{algorithmic}[1]
\item[\textbf{Inputs:}] Per-workload simulation logs $\{\Lambda_i\}$,
         target SLO violation rate $\tau_\text{target}$
\item[\textbf{Configuration:}] Delinquency threshold $k$,
         spinup delay $\delta$, spacing $\sigma_{\min}$
\item[\textbf{Outputs:}]  List $\Gamma$ of promising spinup placement times.

\State $\mathrm{events} \gets$ All query arrival/completion events from simulation logs $\{\Lambda_i\}$
\State $\mathrm{events} \gets \Call{Sort}{\mathrm{events}, \{\mathrm{time}, \text{\textsc{Asc}}}\}$
\State $V_i \gets 0$, $N_i \gets 0$, $D_i \gets \mathrm{false}$ for all $i$
\State $n_D \gets 0$;\quad $\mathrm{inEpoch} \gets \mathrm{false}$;\quad $\Gamma \gets [\,]$
\For{$j = 0$ \textbf{to} $|\mathrm{events}| - 2$}
    \State $e \gets \mathrm{events}[j]$
    \If{$e.\mathrm{type} = \textsc{Start}$}
        \State $V_{e.i} \mathrel{+}= e.v$;\quad $N_{e.i} \mathrel{+}= 1$
    \Else
        \State $V_{e.i} \mathrel{-}= e.v$;\quad $N_{e.i} \mathrel{-}= 1$
    \EndIf
    \State $d' \gets \frac{V_{e.i}}{N_{e.i}} \geq \tau_\text{target}$
    \If{$d' \neq D_{e.i}$}
        \State $n_D \mathrel{+}= (d'\mathbin{?}{+1}:{-1})$;\quad $D_{e.i} \gets d'$
    \EndIf
    \State \textbf{if} $\mathrm{events}[j{+}1].\mathrm{time} = e.\mathrm{time}$:\;\textbf{continue}
    \If{$n_D \geq k$ \textbf{ and not } $\mathrm{inEpoch}$}
        \State $\mathrm{inEpoch} \gets \mathrm{true}$
        \State $t_\mathrm{cand} \gets e.\mathrm{time} - \delta$
    \ElsIf{$n_D < k$ \textbf{ and } $\mathrm{inEpoch}$}
        \State $\mathrm{inEpoch} \gets \mathrm{false}$
        \If{$\{t \in \Gamma \mid |t-(t_\text{cand})| < \sigma_{\min}\} = \emptyset$}
        \State $\Gamma.\Call{Append}{t_\text{cand}}$
        \EndIf
    \EndIf
\EndFor
\If{$\mathrm{inEpoch}$ \textbf{ and } $\{t \in \Gamma \mid |t-(t_\text{cand})| < \sigma_{\min}\} = \emptyset$}
        \State $\Gamma.\Call{Append}{t_\text{cand}}$
\EndIf
\State \Return $\Gamma$

\end{algorithmic}
\end{algorithm}

\begin{figure}
    \centering
    \includegraphics[width=0.8\linewidth]{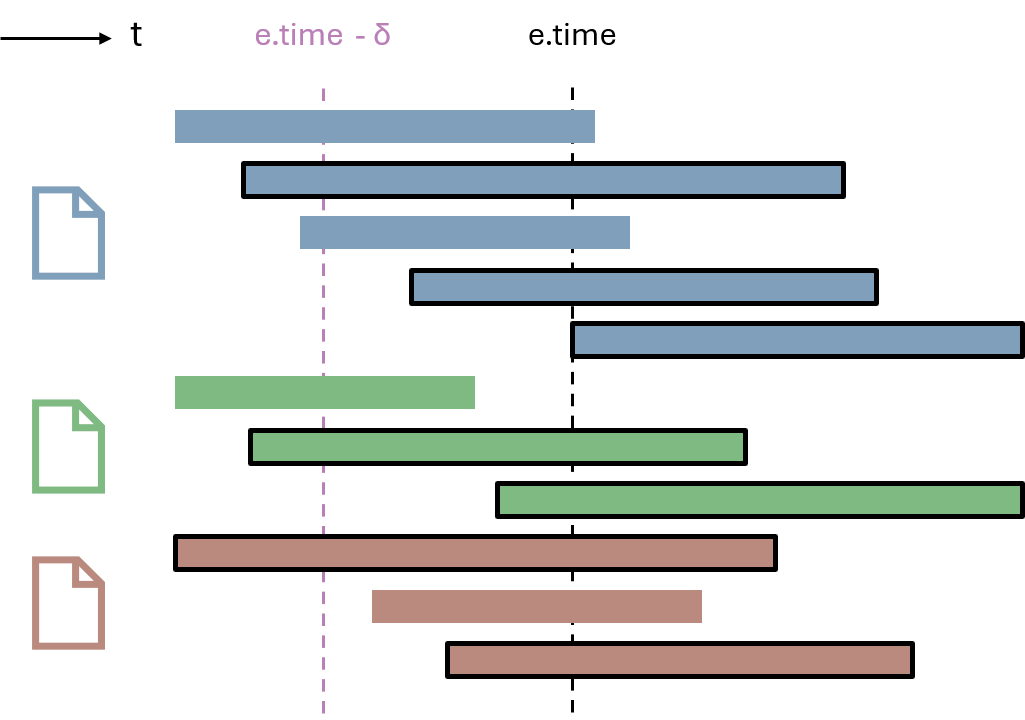}        

    \caption{Candidate spinup time determination. Here, queries with bold outlines violated their SLOs. The indicated time is the first candidate spinup time for $\tau_\text{target}=0.6$ and $k=3$.}
    \label{fig:policy-tuner-spinup}

    \Description{Please refer to the text of Section~\ref{sec:policy-tuner-time} for a Figure walkthrough.}
\end{figure}

Per Section~\ref{sec:policy-tuner-proactive-high}, the \component{Spinup Scheduler} needs to determine promising times for scheduled spinups. Algorithm~\ref{alg:find-spinup-time} achieves this in a single linear scan. It begins by building a unified sorted timeline of query arrival and completion events across all per-training-workload simulation logs for the same execution configuration, pre-gathering per-query SLO violation information (lines~1--2). It then initializes per-workload running-sum accumulators $V_i$, $N_i$, delinquency flags $D_i$, the global delinquent-workload count $n_D$, and the epoch-tracking state (lines~3--4). For each event $e$, the running sums for the affected workload are updated (lines~6--10), and the workload's delinquency flag and $n_D$ are refreshed (lines~11--13). Using running sums means that SLO adherence is evaluated in $O(1)$ per event. Zero-length intervals between simultaneous events are skipped (line~14). When $n_D$ first reaches the threshold $k$, a \emph{congestion epoch} begins and a candidate spinup time is recorded $\delta$ seconds earlier  (lines~15--17); when $n_D$ drops back below $k$, the epoch ends and the candidate spinup time  is appended to $\Gamma$ only if it is at least $\sigma_{\min}$ ahead of the previous candidate (lines~18--21). Any open epoch at the end of the timeline is closed after the loop (lines~22--23). The returned list $\Gamma$ (line 24) is implicitly sorted in ascending order by placement time.

Figure~\ref{fig:policy-tuner-spinup} shows an example with 3 simulation logs, with queries colored blue, green and red respectively. Bold outlines indicate SLO violations.  For $\tau_\text{target}=0.6$ and $k=3$, all 3 workloads have more than $60\%$ of their running queries violating the SLO once the fifth blue query arrives, suggesting a candidate spinup time $\delta$ s earlier.

\sparagraph{Algorithm~\ref{alg:find-spinup-time} Computational Time Complexity}
Let $Q = \sum_i|\Lambda_i|$, so that $|\text{events}| = 2Q$. The event timeline is sorted and then processed once, so that \textsc{FindGoodSpinupTime} is $O(Q\log Q)$.

\subsection{Configuration Tuner}\label{sec:policy-tuner-sweep}

After we determine the scheduled spinups, the \component{Configuration Tuner} optimizes the parameters of the \component{Autoscaler}. It first generates a set of candidate execution configurations using a configurable strategy; supported options include \texttt{grid} (exhaustive cross-product of specified values), \texttt{random} (budget-limited random sample from the grid), \texttt{coordinate\_descent} (iterative per-parameter optimization), and \texttt{adaptive\_batch} (progressive widening with early stopping). It then evaluates all candidate configurations on the training workloads in parallel via \textsc{SimulateBatch} and retains only the top-$k$ with respect to a configurable aggregate metric over the training workloads (e.g. mean). These are re-evaluated (using \textsc{SimulateBatch}) on the validation workloads, and the configuration with the best aggregate validation performance is returned.

\section{Evaluation}\label{sec:evaluation}

We  implemented \thesystem in around 26k lines of Python~\cite{autoslo-repo}. After explaining our evaluation setup (Section~\ref{sec:evaluation-setup}), we will present end-to-end experiments (Section~\ref{sec:evaluation-e2e}) and assess the effectiveness  (Sections~\ref{sec:evaluation-policy-tuner}-~\ref{sec:evaluation-model}) and  efficiency (Section~\ref{sec:evaluation-efficiency}) of each component. In Tables~\ref{tab:evaluation-query-router-effectiveness}--~\ref{tab:evaluation-policy-tuner}, VR denotes the SLO violation rate and bold/underline indicates the best/second-best VR per row.

\subsection{Setup}\label{sec:evaluation-setup}

\subsubsection{Configuration}\label{sec:evaluation-setup-config} We  use a machine with two 20-core 2.10 GHz Intel Xeon Gold 6230 CPUs~\cite{cpu} and 256 GiB of memory, running Linux 6.9.7-arch1-1. From this machine, we use \texttt{psycopg2} and \texttt{boto3} to interact with Amazon Redshift Serverless. For continuity, we will refer to workgroups as  ``clusters'' below. We use the current Amazon Redshift Serverless price for \texttt{us-east-1} ($\$0.375/\text{RPU}/h$), where an RPU is the unit of cluster size. 

\subsubsection{Workloads}\label{sec:evaluation-setup-workloads} All experiments are executed against TPC-DS SF 1000 (1 TB). Several of our experiments use a workload consisting of 3 copies of one query per benchmark query template, for a total of 297 queries, shuffled and issued according to a Poisson process with rate $\lambda$. We will call such workloads \baseline{BaseWorkload}($\lambda$). Other experiments will each describe the workload(s) they use.

\subsubsection{SLOs}\label{sec:evaluation-setup-slos} We run \baseline{BaseWorkload} in a closed loop against a 16-RPU Amazon Redshift Serverless cluster and record the maximum observed latency $\ell_\text{baseline}(t)$ per query template $t$. Below, we let the SLO of each query $q$ of template $t$ be $\kappa \cdot\ell_\text{baseline}(t)$, with higher $\kappa$ indicating looser/easier to satisfy SLOs.


\subsection{End-to-end Effectiveness}\label{sec:evaluation-e2e}

\begin{figure}[]
\centering

    \begin{subfigure}{0.49\linewidth}
        \centering
        \includegraphics[width=\linewidth]{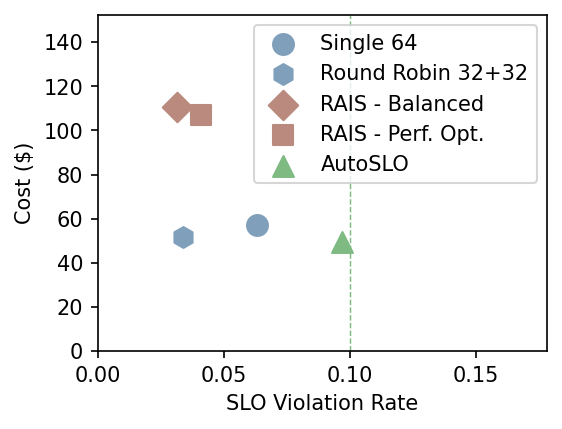}

        \caption{May 27, $\kappa=4$}\label{fig:evaluation-e2e-results-may27-kappa4}
    \end{subfigure}
    \hfill
    \begin{subfigure}{0.49\linewidth}
        \centering
        \includegraphics[width=\linewidth]{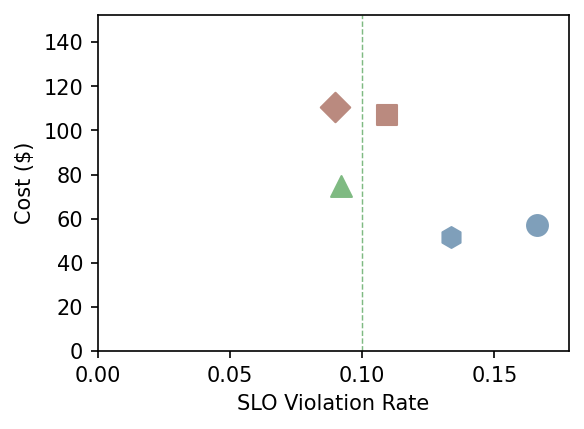}

        \caption{May 27, $\kappa=2$}\label{fig:evaluation-e2e-results-may27-kappa2}
    \end{subfigure}
   
    \begin{subfigure}{0.49\linewidth}
        \centering
        \includegraphics[width=\linewidth]{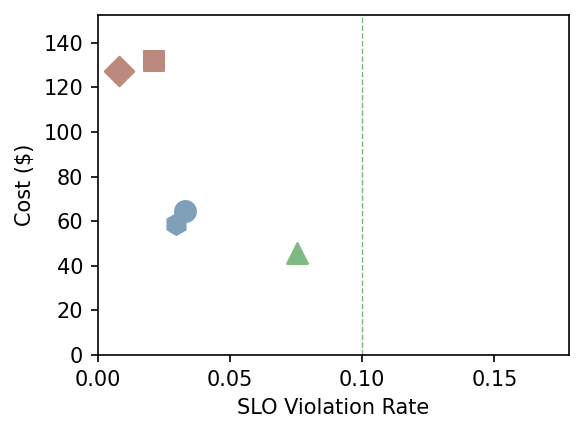}

        \caption{April 15, $\kappa=4$}\label{fig:evaluation-e2e-results-april15-kappa4}
    \end{subfigure}
    \hfill
    \begin{subfigure}{0.49\linewidth}
        \centering
        \includegraphics[width=\linewidth]{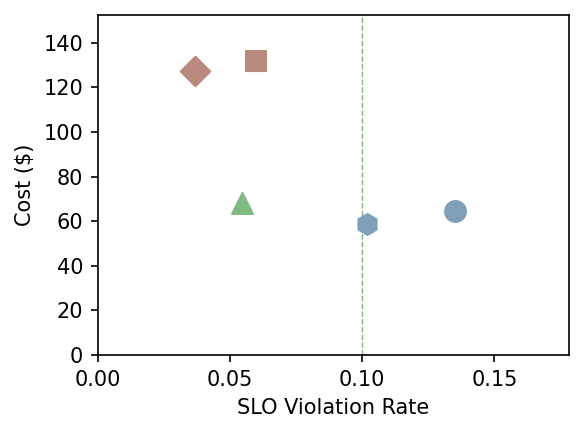}

        \caption{April 15, $\kappa=2$}\label{fig:evaluation-e2e-results-april15-kappa2}
    \end{subfigure}
\hfill

\caption{
        End-to-end performance of \thesystem.}
    \label{fig:evaluation-e2e-results}

    \Description{Please refer to the text of Section~\ref{sec:evaluation-e2e} for a Figure walkthrough.}

\end{figure}

We begin by evaluating \thesystem end-to-end using Redbench~\cite{wehrstein2025redbenchworkloadsynthesiscloud}, which creates realistic workloads based on the real production traces in Redset~\cite{renen2024why}. Redset includes query timing and complexity information for real Amazon Redshift customer workloads, but lacks actual query texts and data. Redbench bridges this gap by mapping each Redset query to appropriate query texts from TPC-DS~\cite{poess2002tpc}. We  obtain an executable version of the SELECT workload faced by a particular Redset cluster (provisioned cluster 157) using Redbench's join matching algorithm. For practical purposes, we compress the interarrival times in the workload by a factor of 6 during simulation/execution, so that each day's workload takes 4 hours to execute. In the \component{Policy Tuner}, this compression is only applied in the \component{Batch Simulator} (i.e. forecasting uses real times).

Based on this workload, we derive 4 experimental scenarios as the cross product of two variables, as shown in Figure~\ref{fig:evaluation-e2e-results}. First, we vary the workload day: we use the last Monday (May 27, 2024) and middle Monday (April 15, 2024) of the workload, to ensure non-overlapping recent workload histories in the \component{Policy Tuner}. Then, we examine two different levels of SLO difficulty, for $\kappa \in \{2, 4\}$. Across scenarios, our SLO violation rate target per Definition~\ref{def:problem} is $\tau_\text{target} = 0.1$, shown as a dashed green line. 

We compare the performance of \thesystem (green triangle), after using the \component{Policy Tuner} over 30 days of past data, against two families of baselines. In blue, we have a pair of naive baselines using a total of 64 RPU each, either in a single cluster (circle) or in two clusters with 32 RPU each, with round-robin query routing (hexagon). In red, we have a single cluster using Redshift Serverless AI-driven Scaling~\cite{nathan2024intelligent}, with the price-performance slider set to either ``balanced'' (position 50, diamond) or ``high performance'' (position 100, square), and a specified maximum of 128 RPU\footnote{We observe that throughout the execution of our workloads, RAIS bills for the full 128 RPU (according to the billing timer of Section~\ref{sec:defs}) regardless of the slider setting.}. 

As evident, \thesystem provides the best performance across scenarios, meeting $\tau_\text{target}$ for all 4 scenarios and reducing cost by a mean of $26.4\%$ compared to the next-best baseline per scenario. Compared to the only other baseline that meets $\tau_\text{target}$ in all 4 scenarios (RAIS - Balanced), \thesystem reduces cost by a mean of $49.6\%$.

Per Section~\ref{sec:introduction}, neither family of baselines allows explicitly specifying SLOs or $\tau_\text{target}$, which translates to ineffective cost-performance tradeoffs.  While the SLOs are loose ($\kappa=4$, Figures~\ref{fig:evaluation-e2e-results-may27-kappa4} and~\ref{fig:evaluation-e2e-results-april15-kappa4}), all baselines meet $\tau_\text{target}$ but remain unaware of it, spending additional resources to reduce latency in a regime where it no longer matters. When the SLOs tighten ($\kappa=2$, Figures~\ref{fig:evaluation-e2e-results-may27-kappa2} and~\ref{fig:evaluation-e2e-results-april15-kappa2}), no baseline can be informed, since they are SLO-unaware; they each act the same as before, leading to higher SLO violation rates, with the naive baselines all missing $\tau_\text{target}$. \thesystem instead adapts, increasing spending just enough to meet $\tau_\text{target}$.

\begin{findings}{End-to-end SLO Adherence}
    \thesystem can cost-effectively maintain latency SLOs on realistic workloads at the specified target level, reducing cost by a mean of $49.2\%$ against the cheapest RAIS baseline per scenario.
\end{findings}


\subsection{Latency Predictor (Iconq+)}\label{sec:evaluation-model}

\subsubsection{Variants}\label{sec:evaluation-model-variants} 

To assess our \component{Latency Predictor}, we compare four model variants in an ablation study:  \baseline{Iconq} includes no changes~\cite{wu2025improving}; \baseline{+Size} enhances the interaction feature vectors with cluster size information (Section~\ref{sec:model-size}); \baseline{+Censored} additionally trains the model on censored observations sampled with probability $0.5\%$ (Section~\ref{sec:model-censored}); finally, \baseline{Iconq+} also supports incremental predictions (Section~\ref{sec:model-incremental}).

\subsubsection{Training Details}\label{sec:evaluation-model-training}

We train each variant on data obtained by executing \baseline{BaseWorkload} for $\lambda \in [0.05, 0.1, 0.2, 0.5, 1.0]$ on each of 4, 8, 16 and 32 RPU. On each cluster size, we also execute four "phased" versions of \baseline{BaseWorkload}, alternating between $x$ queries at $\lambda=0.05$ and 10 queries at $\lambda_{high}$, for $x\in\{40, 15\}$ and $\lambda_{high}\in\{0.2, 0.5\}$. For practicality, we impose an one-hour per-query timeout when executing these workload runs.
For all four variants, we also perform two training stability modifications  compared to the original Iconq paper~\cite{wu2025improving}. First, we only train the underlying Stage~\cite{wu2024stage} model (used to derive neighbor-unaware latency predictions included  in interaction feature vectors as query complexity proxies) on isolated query executions, i.e., executions with no neighbors. This makes the resulting predictions better reflect intrinsic query complexity rather than interference from concurrent queries. Second, we interpret data labels and model latency predictions in \emph{logarithmic space}, guaranteeing positivity after re-exponentiation and aligning naturally with multiplicative error. This means we no longer need to heavily penalize negative/tiny predictions in the loss function, which can now use a simpler Q-error-based objective.

\subsubsection{Results}\label{sec:evaluation-model-results} 

\begin{figure}
    \centering

\begin{subfigure}[t]{0.6\linewidth}
    \centering
        \includegraphics[width=\linewidth]{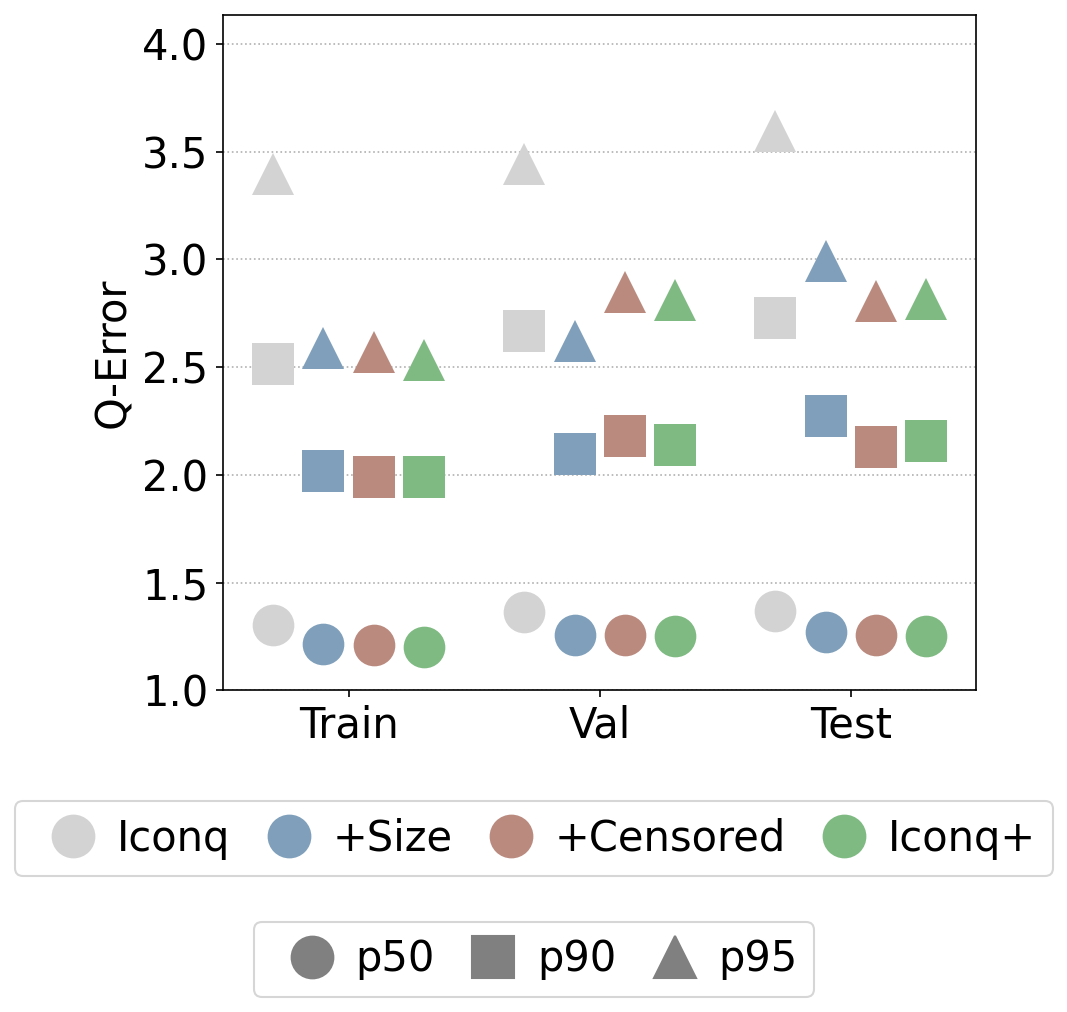}

        \caption{Accuracy}
        \label{fig:evaluation-model-accuracy}
    \end{subfigure}
    
    \begin{subfigure}[t]{0.8\linewidth}
        \includegraphics[width=\linewidth]{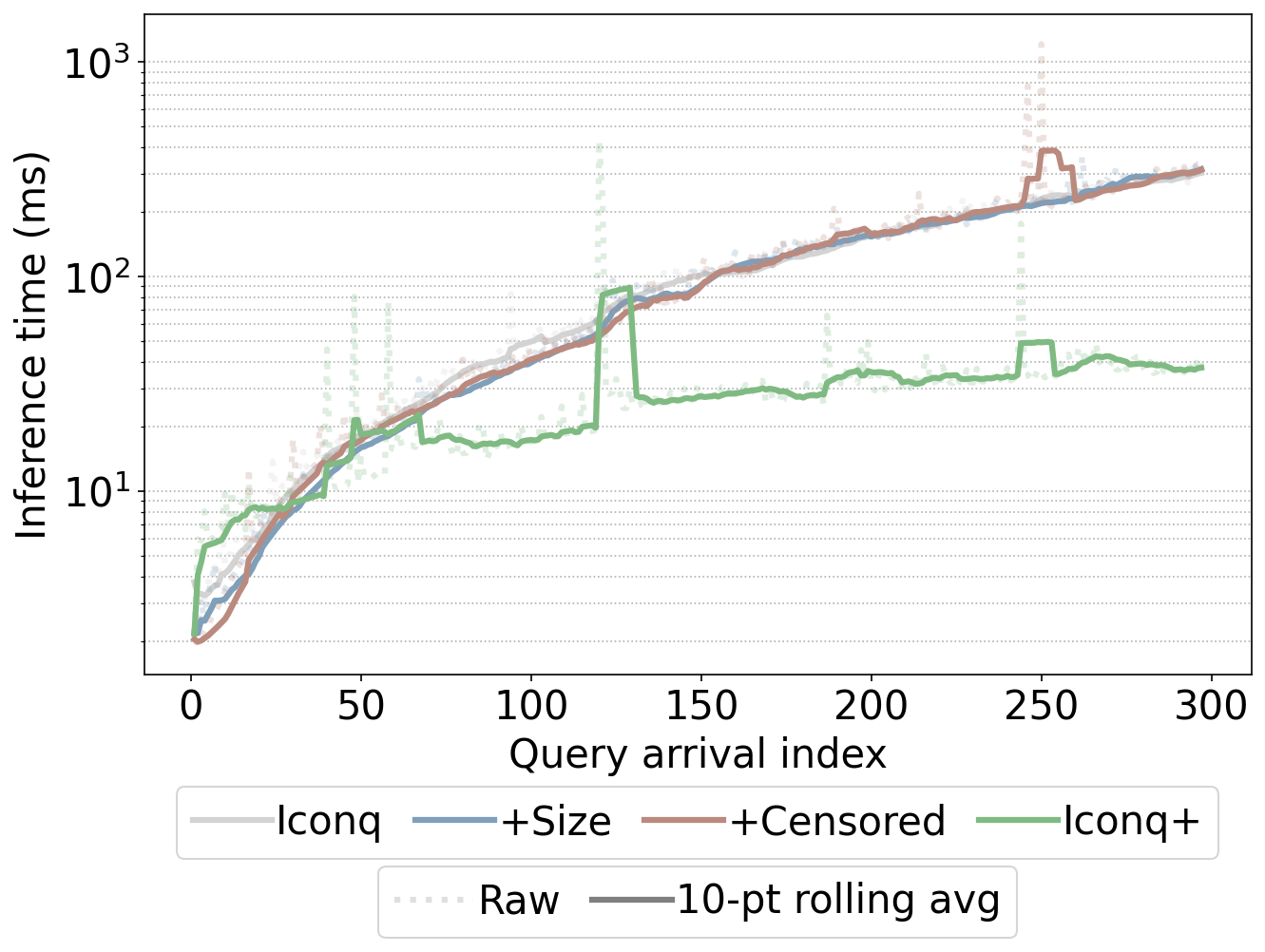}

        \caption{Efficiency}
        \label{fig:evaluation-model-efficiency}
    \end{subfigure}

    \caption{Iconq+ delivers superior accuracy and efficiency.}
    \label{fig:evaluation-model}

    \Description{Please refer to the text of Section~\ref{sec:evaluation-model-results} for a Figure walkthrough.}

\end{figure}

Figure~\ref{fig:evaluation-model-accuracy} evaluates the accuracy of each model. It shows the 50th, 90th and 95th percentile of the Q-error achieved by each variant on each of its folds. As evident, adding the cluster-size-related features in \baseline{+Size} offers noticeable benefits on the test set, in addition to being necessary for cross-cluster-size comparisons when routing. \baseline{+Censored} extends these benefits further, especially at p90 and p95, with \baseline{Iconq+} achieving similar accuracy.

Figure~\ref{fig:evaluation-model-efficiency} concerns the efficiency of the different variants on \baseline{BaseWorkload} with $\lambda=0.1$. For each query arrival, we predict its own latency and update the latency predictions of previous queries, assuming no query has completed yet. This is  not reflective of a real execution of this particular workload, but this experiment is only stress-testing total inference latency per arrival under increasing load. As evident, inference time grows with increasing congestion, but \baseline{Iconq+} can use incremental inference to sustain sub-40 ms inference latency. In contrast, every other variant requires around 300 ms by the end of the workload.

\begin{findings}{Latency Predictor (Iconq+)}
    The adaptations of Iconq+ are effective: adding cluster size features and  censored observations noticeably prediction accuracy, while incremental inference keeps inference time manageable.
\end{findings}

\subsection{Query Router}\label{sec:evaluation-query-router}

We start by assessing the performance of the \component{Query Router}. We compare three query routing approaches. \baseline{Round-Robin} cycles among the active clusters, as a naive baseline. \baseline{Stage + Cost Opt.} uses the underlying Stage~\cite{wu2024stage} model of our trained Iconq+ instance to derive concurrency-unaware latency predictions on each active cluster and then routes each query to a cluster where its SLO will be met, breaking ties by cost. Such concurrency-agnostic routing mirrors the approach of published descriptions from cloud vendors~\cite{saxena2023auto, nathan2024intelligent}. \baseline{Ours (Iconq+ + Cost Opt.)} represents our approach as described in Section~\ref{sec:query-router}, where the concurrency-aware Iconq+ model is used and the risk to the SLOs of running queries is also assessed.

We evaluate each approach in 8 scenarios, derived as follows: (i) we run \baseline{BaseWorkload} with either $\lambda=0.1$ or $\lambda=0.2$; (ii) we define SLOs using either $\kappa=4$ or $\kappa=2$; and (iii)  we either route between two clusters with 16 RPU each ($\mathcal{C} = [16,16]$), or between one cluster each with 32 RPU and 16 RPU ($\mathcal{C} = [32,16]$).

Table~\ref{tab:evaluation-query-router-effectiveness} shows the results. Query latency prediction with Stage already outperforms \baseline{Round-Robin}, reducing the SLO violation rate by a mean of $24.3\%$. It also reduces cost by a mean of $19.3\%$, since the \component{Query Router} also considers cost. However, using  Iconq+ can improve even further over \baseline{Round-Robin}, achieving a mean SLO violation rate reduction of $47.8\%$ while still lowering cost.

\begin{table}[t]
\centering
\small
\setlength{\tabcolsep}{1.75pt}

\caption{Query Router evaluation.}
\label{tab:evaluation-query-router-effectiveness}

\begin{tabular}{c c c || c c | c c | c c}
\toprule
$\lambda$ & $\kappa$ & C & \multicolumn{2}{c|}{Round Robin} & \multicolumn{2}{c|}{Stage +} & \multicolumn{2}{c}{Iconq +} \\
 &  &  &  &  & \multicolumn{2}{c|}{Cost.Opt.} & \multicolumn{2}{c}{Cost.Opt.} \\

 &  &  & VR & Cost & VR & Cost & VR & Cost \\
\hline
0.1 & 4 & [32, 16] & 0.0269 & \$15.78 & \textbf{0.0135} & \$11.01 & \underline{0.0168} & \$11.17 \\
0.1 & 4 & [16, 16] & 0.0976 & \$10.40 & \underline{0.0370} & \$7.99 & \textbf{0.0303} & \$8.91 \\
0.1 & 2 & [32, 16] & 0.0842 & \$15.78 & \underline{0.0808} & \$10.14 & \textbf{0.0505} & \$12.28 \\
0.1 & 2 & [16, 16] & 0.2088 & \$10.57 & \underline{0.1414} & \$7.43 & \textbf{0.1246} & \$8.77 \\
0.2 & 4 & [32, 16] & 0.1414 & \$7.92 & \underline{0.0707} & \$7.56 & \textbf{0.0202} & \$7.60 \\
0.2 & 4 & [16, 16] & 0.3064 & \$5.41 & \underline{0.2761} & \$4.81 & \textbf{0.1919} & \$5.29 \\
0.2 & 2 & [32, 16] & 0.1919 & \$8.13 & \underline{0.1919} & \$6.92 & \textbf{0.1044} & \$7.71 \\
0.2 & 2 & [16, 16] & \underline{0.4512} & \$5.48 & 0.5118 & \$5.20 & \textbf{0.3300} & \$4.98 \\
\hline
\multicolumn{3}{c||}{\textit{Mean~$\Delta$}} & --- & --- & \underline{-24.3\%} & -19.3\% & \textbf{-47.8\%} & -12.9\% \\
\bottomrule
\end{tabular}

\end{table}

\begin{findings}{Query Router}
    Our \component{Query Router} outperforms the alternatives, reducing SLO violation rate by a mean of $47.8\%$ and cost by a mean of $12.9\%$.
\end{findings}


\subsection{Autoscaler -- Spinup Size Selector}\label{sec:evaluation-autoscaler-size-selection}

\begin{table}[t]
\centering
\small
\setlength{\tabcolsep}{1.75pt}
\caption{Spinup Size Selector steady-state evaluation.} \label{tab:evaluation-autoscaler-size-selection}

\begin{tabular}{c c c || c c | c c | c c | c c}
\toprule
$\lambda$ & $\kappa$ & C & \multicolumn{2}{c|}{No-Op} & \multicolumn{2}{c|}{Horizontal} & \multicolumn{2}{c|}{Vertical} & \multicolumn{2}{c}{Add Cluster} \\
 &  &  &  &  &  &  & \multicolumn{2}{c|}{w/Ours} & \multicolumn{2}{c}{w/Ours} \\

 &  &  & VR & Cost & VR & Cost & VR & Cost & VR & Cost \\
\hline
0.1 & 4 & [8] & 0.9833 & \$2.81 & 0.8167 & \$1.99 & \underline{0.2000} & \$1.79 & \textbf{0.0167} & \$2.16 \\
0.1 & 4 & [16] & 0.1000 & \$1.65 & \textbf{0.0000} & \$2.67 & 0.0833 & \$1.59 & \textbf{0.0000} & \$2.35 \\
0.1 & 2 & [8] & 0.9833 & \$2.86 & 1.0000 & \$2.04 & \textbf{0.0167} & \$3.40 & \underline{0.1000} & \$2.09 \\
0.1 & 2 & [16] & 0.3667 & \$1.74 & \underline{0.0667} & \$3.14 & \textbf{0.0167} & \$3.40 & 0.0833 & \$2.81 \\
0.2 & 4 & [8] & 1.0000 & \$3.27 & 1.0000 & \$3.55 & \textbf{0.0333} & \$3.11 & \textbf{0.0333} & \$2.88 \\
0.2 & 4 & [16] & 0.8333 & \$2.21 & 0.1167 & \$1.83 & \underline{0.0333} & \$2.14 & \textbf{0.0000} & \$2.71 \\
0.2 & 2 & [8] & 1.0000 & \$3.31 & 0.9500 & \$3.34 & \textbf{0.0667} & \$3.12 & \textbf{0.0667} & \$3.15 \\
0.2 & 2 & [16] & 0.9000 & \$2.05 & 0.4167 & \$2.26 & \underline{0.0667} & \$2.29 & \textbf{0.0500} & \$2.75 \\
\hline
\multicolumn{3}{c||}{\textit{Mean~$\Delta$}} & --- & --- & -42.7\% & +11.0\% & \underline{-83.6\%} & +9.1\% & \textbf{-93.7\%} & +11.8\% \\
\bottomrule
\end{tabular}

\end{table}

We next focus on the cluster sizes selected by the \component{Scaling Controller} within the \component{Autoscaler}. We compare four approaches to modifying the active cluster set. \baseline{No-Op} makes no changes to the active clusters. \baseline{Horizontal} spins up an additional cluster of the same size as the  existing one, using our \component{Query Router} to route between the two. \baseline{Add Cluster w/ Ours} represents our approach described in Section~\ref{sec:autoscaler}. \baseline{Vertical w/ Ours} uses the simulation-based decision-making of our method, but instead of considering what single cluster to \emph{add}, it considers what single cluster to \emph{have} (i.e. it stop routing to the existing cluster once the new one becomes available). 

We evaluate each approach in 8 scenarios, derived as in Section~\ref{sec:evaluation-query-router}, but we start with only a single cluster of size either 8 or 16 RPU. We execute $20\%$ of the workload queries, at which point we \emph{forcibly trigger} autoscaling. We then let the following $60\%$ of the workload elapse, for scaling to take effect and any congestion-affected queries around the autoscaling point to drain. We report the  SLO violation rate and cost over the last $20\%$ of the workload, once a steady state has been reached.

The results are presented in Table~\ref{tab:evaluation-autoscaler-size-selection}. The cluster size selected by the \component{Spinup Size Selector} outperforms horizontal scaling: \baseline{Add Cluster w/ Ours} achieves a mean SLO violation rate reduction of $93.7\%$, compared to $42.7\%$ for \baseline{Horizontal}, while the two methods impact cost similarly. Relying on our algorithm for vertical scaling is also promising, with \baseline{Vertical w/ Ours} achieving a mean SLO violation rate reduction of $83.6\%$. However, \baseline{Add Cluster w/ Ours} performs best on average,  making effective use of the pre-existing cluster.

\begin{findings}{Autoscaler -- Spinup Size Selector}
    The \component{Spinup Size Selector} recommends appropriate cluster sizes. Our approach, where we use the new cluster alongside the existing one, achieves a mean SLO violation rate reduction of $93.7\%$.
\end{findings}


\subsection{Autoscaler -- Spinup Trigger}\label{sec:evaluation-autoscaler-trigger} 

In the previous experiment, we focused on \emph{which} cluster the \component{Autoscaler} spins up, once triggered; we now also consider \emph{when} it is triggered. We compare four approaches to triggering autoscaling, all of which then use the standard \component{Spinup Size Selector}. \baseline{No-Op} is never triggered. \baseline{Queue@32} is triggered whenever there are 32 or more outstanding queries, approximating queuing-based autoscaling schemes present in Auto-WLM~\cite{saxena2023auto} and RAIS~\cite{nathan2024intelligent}. \baseline{Observed} is similar to Algorithm~\ref{alg:autoscaler-spinup-trigger} but without $R$ (line 5); it only acts on \emph{observed} latencies. Finally, \baseline{Ours} represents our approach described in Section~\ref{sec:autoscaler-triggers-spinup}, which augments \baseline{Observed} with the projected SLO violation status of currently-running queries.

We use the same 8 experimental scenarios as Section~\ref{sec:evaluation-autoscaler-size-selection}. 
Since each approach can decide when and what to spin up, we compare SLO violation rate and cost throughout the whole workload. As shown in Table~\ref{tab:evaluation-autoscaler-trigger}, \baseline{Ours} clearly outperforms the alternatives on average, reducing the SLO violation rate by a mean of $54.6\%$.

\begin{table}
\centering
\small
\setlength{\tabcolsep}{1.75pt}
\caption{Autoscaler Spinup Trigger evaluation.}

\label{tab:evaluation-autoscaler-trigger}

\begin{tabular}{c c c || c c | c c | c c | c c}
\toprule
$\lambda$ & $\kappa$ & C & \multicolumn{2}{c|}{No-Op} & \multicolumn{2}{c|}{Queue@32} & \multicolumn{2}{c|}{Observed} & \multicolumn{2}{c}{Ours} \\
 &  &  & VR & Cost & VR & Cost & VR & Cost & VR & Cost \\
\hline
0.1 & 4 & [8] & 0.9697 & \$4.56 & 0.4276 & \$6.10 & \underline{0.4007} & \$5.88 & \textbf{0.3199} & \$7.31 \\
0.1 & 4 & [16] & 0.2896 & \$5.31 & 0.3064 & \$5.09 & \textbf{0.2222} & \$7.18 & \underline{0.2626} & \$7.83 \\
0.1 & 2 & [8] & 0.9865 & \$4.71 & \underline{0.4242} & \$8.67 & 0.8013 & \$6.38 & \textbf{0.2458} & \$11.10 \\
0.1 & 2 & [16] & 0.4882 & \$5.24 & 0.4916 & \$5.25 & \underline{0.2727} & \$8.68 & \textbf{0.2660} & \$8.74 \\
0.2 & 4 & [8] & 0.9899 & \$4.13 & \textbf{0.2896} & \$5.54 & 0.9327 & \$5.04 & \underline{0.3098} & \$5.19 \\
0.2 & 4 & [16] & 0.8114 & \$3.71 & 0.4141 & \$5.72 & \textbf{0.3434} & \$5.91 & \underline{0.4007} & \$4.95 \\
0.2 & 2 & [8] & 0.9966 & \$4.16 & \underline{0.4680} & \$5.37 & 0.5791 & \$5.91 & \textbf{0.4040} & \$6.02 \\
0.2 & 2 & [16] & 0.9495 & \$3.65 & 0.4882 & \$5.45 & \underline{0.4680} & \$5.77 & \textbf{0.3670} & \$7.27 \\
\hline
\multicolumn{3}{c||}{\textit{Mean~$\Delta$}} & --- & --- & \underline{-41.0\%} & +35.1\% & -37.6\% & +43.3\% & \textbf{-54.6\%} & +64.1\% \\
\bottomrule
\end{tabular}

\end{table}

\begin{findings}{Autoscaler -- Spinup Trigger}
    Our \component{Autoscaler} spinup trigger, combining observed and projected SLO violations, outperforms the alternatives and delivers a mean SLO violation rate reduction of $54.6\%$.
\end{findings}


\subsection{Policy Tuner}\label{sec:evaluation-policy-tuner}

\begin{table}[t]
\centering
\small
\setlength{\tabcolsep}{1.75pt}
\caption{Policy Tuner evaluation.}\label{tab:evaluation-policy-tuner}

\begin{tabular}{c c || c c | c c | c c | c c}
\toprule
Day & $\kappa$  & \multicolumn{2}{c|}{No Past Data} & \multicolumn{2}{c|}{Past 1 Day} & \multicolumn{2}{c|}{Past 7 Days} & \multicolumn{2}{c}{Past 30 Days} \\
 &  &   VR & Cost & VR & Cost & VR & Cost & VR & Cost \\
\hline
5/27  & 4 & 0.1997 & \$41.47 & 0.1149 & \$46.14 & \textbf{0.0639} & \$51.75 & \underline{0.0970} & \$49.25 \\
5/27 &  2  & 0.2162 & \$61.72 & \underline{0.1092} & \$71.39 & 0.1293 & \$68.38 & \textbf{0.0920} & \$74.89 \\
4/15  & 4 & 0.0833 & \$42.64 & \textbf{0.0664} & \$44.15 & 0.0848 & \$45.65 & \underline{0.0752} & \$45.82 \\
4/15& 2   & 0.1600 & \$59.80 & \textbf{0.0546} & \$69.78 & 0.0575 & \$68.11 & \textbf{0.0546} & \$68.01 \\
\hline
\multicolumn{2}{c||}{\textit{Mean~$\Delta$}} & --- & --- & \underline{-44.6\%} & +11.8\% & -42.6\% & +14.1\% & \textbf{-46.1\%} & +15.3\% \\
\bottomrule
\end{tabular}
\end{table}

We next evaluate the impact of the \component{Policy Tuner} on the downstream performance of the execution configuration it produces. In particular, for the day/$\kappa$ scenarios used in  Section~\ref{sec:evaluation-e2e}, we use the \baseline{Policy Tuner} to optimize a configuration on 0, 1, 7 or 30 days of past data. We then use each optimized configuration to execute the target workload. As seen in Table~\ref{tab:evaluation-policy-tuner}, the \component{Policy Tuner} can leverage even a single day of workload history to reduce the SLO violation rate by a mean of $44.6\%$, increasing cost by a mean of just $11.8\%$. With access to 30 days of history, the mean SLO violation rate reduction further improves to $46.1\%$, but most of the tuning benefit can be already reaped with minimal history.

\begin{findings}{Policy Tuner}
    The Policy Tuner can efficiently produce better execution configurations, reducing the SLO violation rate by a mean of $44.6\%$ with access to a single day of workload history.
\end{findings}


\subsection{Efficiency}\label{sec:evaluation-efficiency}

\begin{figure}[t]
    \centering

    \begin{subfigure}[t]{0.49\linewidth}
        \centering
        \includegraphics[width=\linewidth]{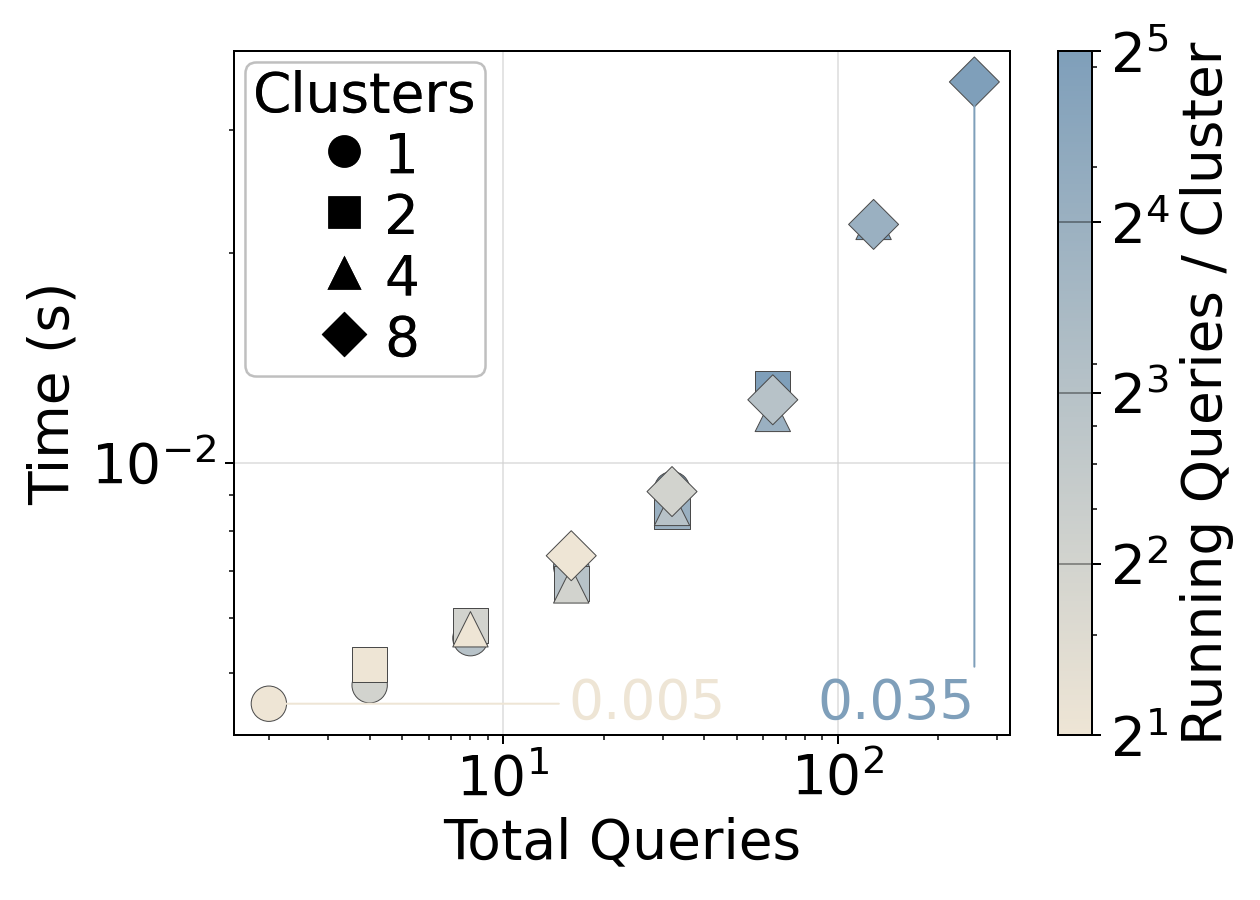}
        \caption{\textsc{Route}}
        \label{fig:evaluation-efficiency-query-router}
    \end{subfigure}
    \hfill
    \begin{subfigure}[t]{0.49\linewidth}
        \centering
        \includegraphics[width=\linewidth]{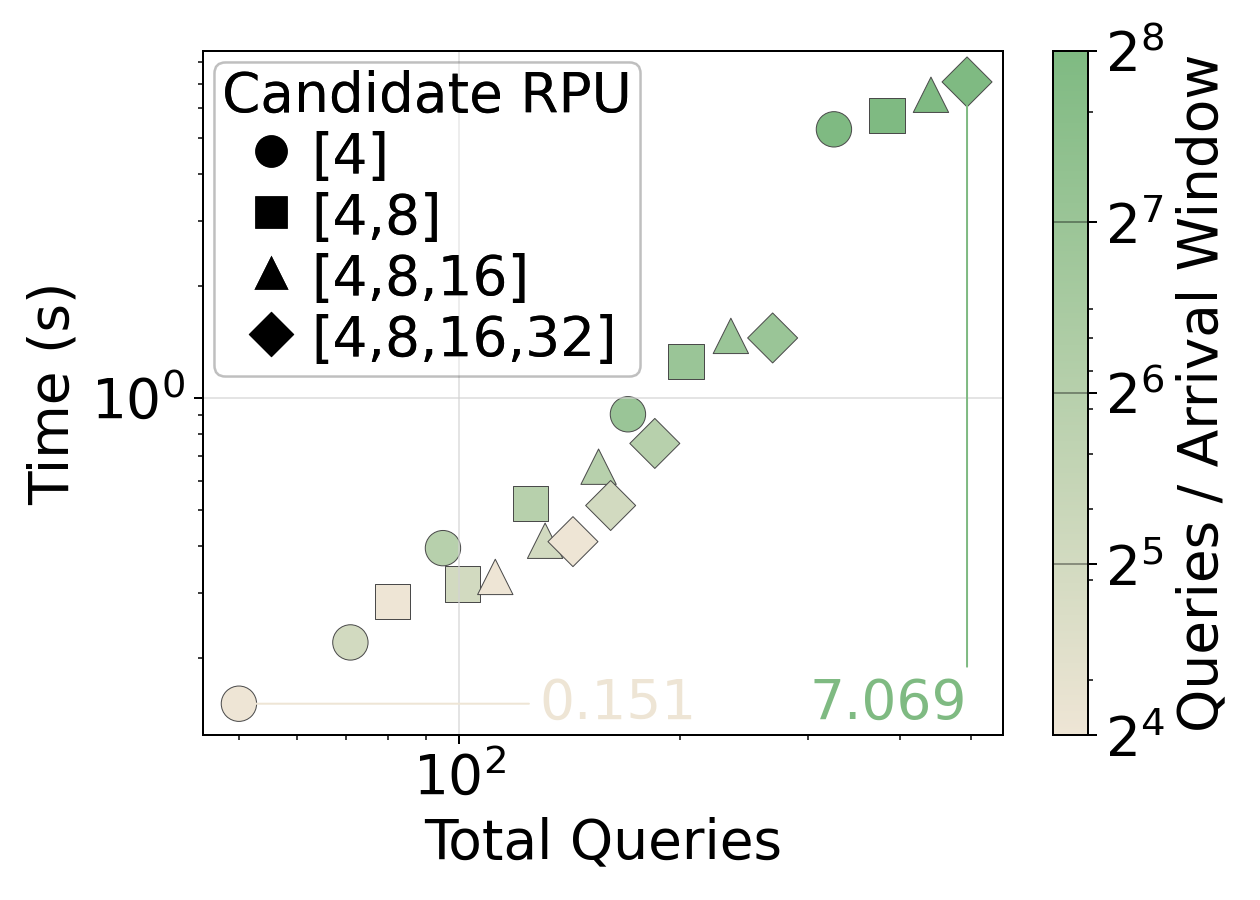}
        \caption{\textsc{FindBestSpinupSize}}
        \label{fig:evaluation-efficiency-autoscaler}
    \end{subfigure}

    \begin{subfigure}[t]{0.49\linewidth}
        \centering
        \includegraphics[width=\linewidth]{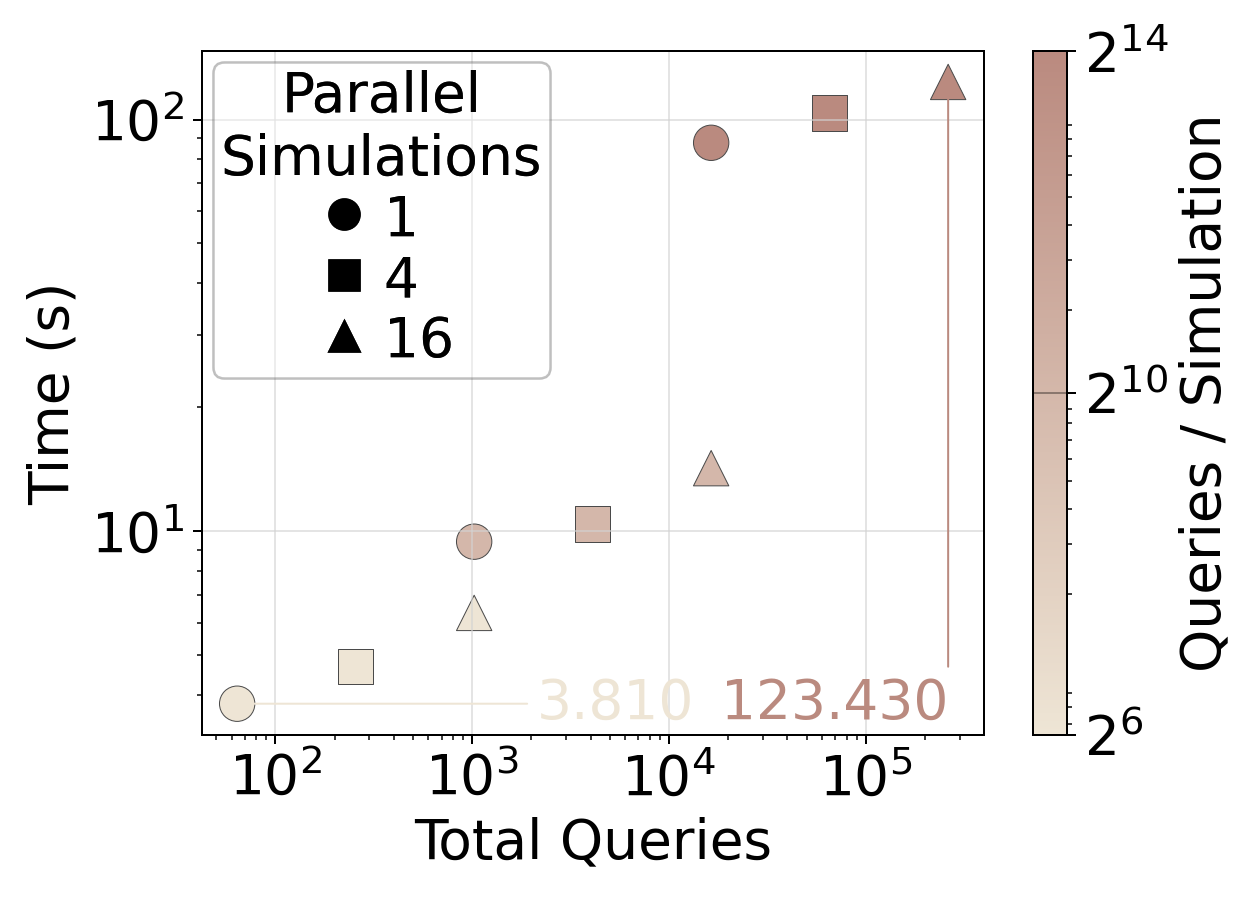}

        \caption{\textsc{SimulateBatch}}
        \label{fig:evaluation-efficiency-simulate-batch}
    \end{subfigure}
    \hfill
    \begin{subfigure}[t]{0.49\linewidth}
        \centering
        \includegraphics[width=\linewidth]{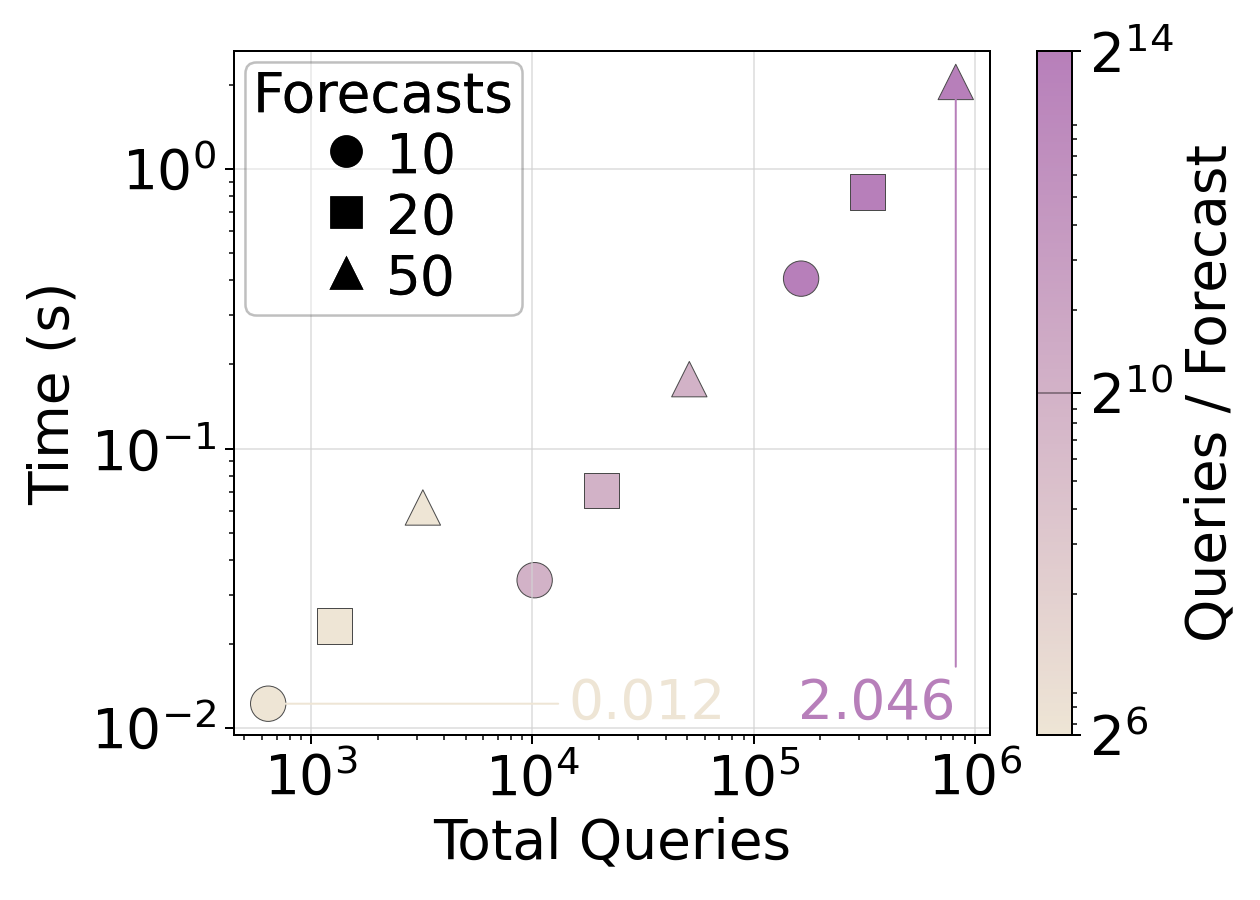}

        \caption{\textsc{FindGoodSpinupTime}}
        \label{fig:evaluation-efficiency-find-good-spinup-time}
    \end{subfigure}

    \caption{Efficiency evaluation of key \thesystem algorithms.}
    \label{fig:evaluation-efficiency}

    \Description{Please refer to the text of Section~\ref{sec:evaluation-efficiency} for a Figure walkthrough.}

\end{figure}

We close with efficiency micro-experiments. In Figure~\ref{fig:evaluation-efficiency}, each point represents the median execution time across 10 repetitions.

\subsubsection{Query Router Efficiency}\label{sec:evaluation-efficiency-query-router} The \component{Query Router} is invoked in the critical path of each query, so it needs to be highly efficient. We assess this by sweeping two variables: the number of active clusters and the number of running queries per cluster. As evident from Figure~\ref{fig:evaluation-efficiency-query-router}, the \component{Query Router} achieves a decision-making time under 10 ms in most scenarios, with a maximum of only 35 ms.

\subsubsection{Autoscaler Efficiency}\label{sec:evaluation-efficiency-autoscaler} 
The \component{Autoscaler} runs in a background thread, as described in Section~\ref{sec:autoscaler}. Still, it must be reasonably efficient for the scaling decision to remain timely. We assess its efficiency by sweeping two variables: the number of candidate cluster sizes and the arrival rate of queries in the \component{Arrival Window}. As seen in Figure~\ref{fig:evaluation-efficiency-autoscaler}, its decision-making time is around 7 s even in the most challenging case, providing fast responsiveness for reactive scaling.

\subsubsection{Policy Tuner Efficiency}\label{sec:evaluation-efficiency-policy-tuner}

The \component{Policy Tuner} is invoked periodically, as described in Section~\ref{sec:policy-tuner}, so its overhead can be somewhat higher. Still, it must not consume excessive resources that could be otherwise used for workload execution. Aside from bookkeeping, two expensive functions are used by the \component{Spinup Scheduler} and the \component{Configuration Tuner}: \textsc{SimulateBatch} and \textsc{FindGoodSpinupTime}. 

For \textsc{SimulateBatch}, we sweep two variables: the number of queries in each simulated workload and the number of workloads simulated in parallel. As seen in Figure~\ref{fig:evaluation-efficiency-simulate-batch}, batch simulation parallelizes well and finishes in around 2 minutes even with over 15,000 queries, well within the performance requirements of an infrequent, background \component{Policy Tuner} invocation.
For \textsc{FindGoodSpinupTime}, we sweep the number of queries in each forecasted workload and the number of forecasts considered. As Figure~\ref{fig:evaluation-efficiency-find-good-spinup-time} shows, the algorithm completes in slightly over 2 s in the worst case.

\begin{findings}{Efficiency}
    Each critical algorithm of \thesystem is efficient, meeting the latency requirements of its operating timescale.
\end{findings}

\section{Related Work}\label{sec:related-work}

\begin{table}[]
    \setlength{\tabcolsep}{1.8pt}
    \caption{Coverage of related work.}
    \label{tab:related-work}
    
    \begin{tabular}{l|c|c|c|c}
    \multicolumn{5}{l}{\textit{Section~\ref{sec:related-work-data}: SLOs for Cloud Data Systems}}\\
    \toprule
        \textbf{Category} &\textbf{Has} & \textbf{Plans}& \textbf{Adjusts}& \textbf{Reacts}  \\
        &\textbf{SLOs}&offline&online&when routing \\
        
        \hhline{=|=|=|=|=}
        \textbf{Only offline planning}&\yes&\yes&\no&\no\\
        \cite{zhang2021restune, marcus2016wisedb}&&&&\\
        \hline
        \textbf{Only online adjustment}&\yes&\no&\yes&\no\\
        \cite{das2016automated, curino2011workload,liu2013pmax, lang2014towards, arora2023flexible, kraska2023check}&&&&\\
        \cite{marcus2017releasing, narasayya2013sqlvm, narasayya2013a, ortiz2018slaorchestrator, ortiz2015changing, ortiz2016perfenforce}&&&&\\
        \cite{xiong2011intelligent, yu2024blueprinting,zhao2013framework, taft2016step, sakr2012sla}&&&&\\
        \hline
        \textbf{No explicit SLOs}&\no&\yes&\yes&\no\\
        \cite{saxena2023auto, nathan2024intelligent, wu2024stage}&&&&\\
        \hline
        \textbf{Fixed, no interference}&\yes&\no&\no&\no\\
        \cite{chi2011sla, chi2011icbs, xiong2011activesla, leitner2012cost, chi2013distribution}&&&&\\
        \hline
        \textbf{Fixed, no SLOs}&\no&\no&\no&\yes\\
        \cite{wu2025improving, ahmad2008qshuffler, rohm2001cache, sabek2022lsched, huang2024laser}&&&&\\
    \bottomrule
    \multicolumn{5}{l}{ }\\
    \multicolumn{5}{l}{\textit{Section~\ref{sec:related-work-other}: SLOs for Other Cloud Systems}}\\
    \toprule
        \textbf{Domain} &  \multicolumn{4}{c}{\textbf{Unlike queries}} \\ 
         &  \multicolumn{4}{c}{\textbf{the scheduled tasks have...}} \\ 
          \hhline{=|====}

        \textbf{Serverless Computing} & \multicolumn{4}{c}{Opaque, function-level behavior}\\
        \cite{kaffes2022hermod, fuerst2021faascache, abdi2023palette}&\multicolumn{4}{l}{ }\\
        \hline
        \textbf{Application Placement} & \multicolumn{4}{c}{Longer lifetimes, } \\
        \cite{delimitrou2013paragon, delimitrou2014quasar,  chen2019parties, romero2018mage, lo2015heracles,patel2020clite}&\multicolumn{4}{c}{SLOs over inbound requests}\\\hline
        \textbf{ML Model Serving} &\multicolumn{4}{c}{Similar computational demands}\\
        \cite{mendoza2021interference, chen2025multiplexing, kossmann2024cascadeserve}&\multicolumn{4}{l}{ }\\\hline
        \textbf{LLM Serving} &\multicolumn{4}{c}{Regular structure,}\\
        \cite{chaudhry2025towards,  hong2025sola, hu2024inference, li2025hotprefix, yu2022orca, kwon2023efficient, zhong2024distserve}& \multicolumn{4}{c}{ only prefix-based caching}\\
    \bottomrule
    \end{tabular}
    
\end{table}

\subsection{SLOs for Cloud Data Systems}\label{sec:related-work-data}  As shown in the top part of Table~\ref{tab:related-work}, no prior work on  cloud data systems offers SLOs and exhibits all three \textbf{key desired behaviors}.  

\sparagraph{Only offline planning} Some works \textbf{plan} for the future workload without adjusting resources online or reacting at routing time. ResTune~\cite{zhang2021restune} reduces resource utilization by replaying the workload on a replica and tuning knobs using constrained optimization. WiseDB~\cite{marcus2016wisedb} trains a decision model on pre-selected query templates to derive fixed template-to-cluster routing decisions. 

\sparagraph{Only online adjustment} On the flip-side, some works only \textbf{adjust} resources based on online load, neither reacting to this load at routing time nor planning future resource allocation. Das et al.~\cite{das2016automated} use an adapted version of the token bucket algorithm~\cite{tanenbaum2011computer} to balance the estimated resource demand of each customer. SLAOrchestrator~\cite{ortiz2018slaorchestrator, ortiz2015changing, ortiz2016perfenforce} suggests and enforces per-query latency SLOs over workloads of ad hoc SELECT-PROJECT-JOIN (SPJ) queries, by having users select their desired cost/performance tier among provided options. Most recently, BRAD~\cite{yu2024blueprinting} focuses on how to select, configure and route among \textit{different engines}. An additional number of works focuses on \textit{vendor-side} optimization, packing customer workloads onto machines~\cite{xiong2011intelligent, curino2011workload, narasayya2013sqlvm, narasayya2013a, liu2013pmax, lang2014towards, taft2016step, arora2023flexible}, unlike our focus on optimizing the requirements of each and every workload. Earlier works have also evaluated alternatives like having the user fully define the autoscaling triggers and exact corrective actions~\cite{sakr2012sla, zhao2013framework} or using reinforcement learning~\cite{marcus2017releasing}. 

\sparagraph{No explicit SLOs} The currently published mechanism of Amazon Redshift, as introduced in  Auto-WLM~\cite{saxena2023auto} and refined in RAIS~\cite{nathan2024intelligent}, balances \textbf{adjusting} and \textbf{planning} but does not optimize for explicit SLOs, nor does it \textbf{react} to observed load through concurrency-aware predictions. Instead, autoscaling is triggered by query queue length, with a qualitative price-performance sensitivity slider controlling the aggressiveness. Query routing relies on concurrency-unaware latency predictions, as described in Stage~\cite{wu2024stage}.

\sparagraph{Fixed resources} A final class of systems assumes a fixed amount of compute per customer. Some support SLOs but do not \textbf{react} to query interactions when routing~\cite{chi2011sla, chi2011icbs, xiong2011activesla, leitner2012cost, chi2013distribution}; others \textbf{react} to query-level interactions but do not optimize for SLOs~\cite{wu2025improving, ahmad2008qshuffler, rohm2001cache, sabek2022lsched, huang2024laser}.

\subsection{SLOs for Other Cloud Systems}\label{sec:related-work-other} Per the bottom part of Table~\ref{tab:related-work}, related work on SLOs for other cloud systems can also be instructive, although not directly applicable.

\sparagraph{Serverless Computing} One area of focus is scheduling tasks in serverless function-as-a-service offerings (e.g. AWS Lambda~\cite{aws-lambda}). 
In this domain, re-executing the \emph{same} function on warm resources can be faster, but there is no notion of a cross-function ``cache''. This is unlike database queries, where the same cached data can be useful to many syntactically different queries. Nevertheless, some concerns are similar.
Hermod~\cite{kaffes2022hermod} balances function performance and resource efficiency through an execution-time-agnostic, but cost- and load-aware hybrid policy, based on simulation-derived insights.
FaaSCache~\cite{fuerst2021faascache} casts function keep-alive as a caching problem, reducing cold-start overhead using a variant of the Greedy-Dual~\cite{cherkasova1998improving} policy.
Palette load balancing~\cite{abdi2023palette} tags each invocation with a user-provided locality hint (\emph{color}); scheduling then maintains a map from colors to instances. 
Such works prize generality and treat the computational demands of each function as opaque. However, when dealing with database queries, latency prediction is more tractable and a critical source of scheduling-relevant information. 

\sparagraph{Application Placement} Works like Paragon ~\cite{delimitrou2013paragon}, Quasar~\cite{delimitrou2014quasar} and Mage~\cite{romero2018mage} optimize application placement (called \emph{scheduling} in this literature) in datacenters with heterogeneous platforms and resources. They treat resource allocation and assignment as a recommendation problem, using limited profiling to balance hardware utilization and the desired performance. 
Works like Heracles~\cite{lo2015heracles}, PARTIES~\cite{chen2019parties} and CLITE~\cite{patel2020clite} explore co-locating such latency-sensitive applications on the same resources as ``background'', throughput-focused work.
In this context, SLOs are defined for \emph{user requests}, while the scheduled entity is the \textit{long-running application}. This is different than routing each individual SLO-bound database query.

\sparagraph{ML Model Serving} Another line of work studies resource allocation for cloud-deployed ML models subject to inference latency SLOs.
For example, Mendoza et al.~\cite{mendoza2021interference} discuss an interference-aware model co-location policy, based on a random forest regressor trained on offline model/hardware profiling.
Mudi~\cite{chen2025multiplexing} further attempts to co-locate inference tasks with model training while minimizing interference, by estimating inference latency as a piece-wise linear function of allocated resources.
CascadeServe~\cite{kossmann2024cascadeserve} uses model output quality as an additional optimization knob, optimizing the design and placement of \emph{model cascades} that invoke higher-latency inference for ``harder'' requests only.
These systems exploit the relative uniformity of inference requests to a given model, whereas analytical queries can differ substantially in operators, data access patterns, and resource demands

\sparagraph{LLM Serving} LLM serving introduces more per-request variability and state management (e.g. KV-cache entries) compared to traditional model serving. 
Several works deal with these issues for individual requests: 
Orca~\cite{yu2022orca} proposed dealing with variable-length decode phases using iteration-level scheduling and selective batching.
PagedAttention~\cite{kwon2023efficient} uses OS-inspired techniques to reduce KV-cache fragmentation and improve throughput. 
DistServe~\cite{zhong2024distserve} and TetriInfer~\cite{hu2024inference} explore prefill-decode disaggregation, so that the resource allocation for each phase can be separately optimized.
SOLA~\cite{hong2025sola} seeks to balance the different SLOs of prefill and decode through state-aware scheduling.
HotPrefix~\cite{li2025hotprefix} addresses the increasing prevalence of requests with the same prefix (e.g. system prompt) by tracking and leveraging the hotness of each prefix while managing the KV-cache.
More recent works look at optimizing compound AI systems built around LLMs. Murakkab~\cite{chaudhry2025towards}  proposes a declarative abstraction that enables under-the-hood efficiency-driven optimizations. 
However, LLM requests still have a comparatively regular structure (centered on prefill/decode) and only prefix-based caching. Database queries have more diverse plans and richer forms of data reuse, leading to a much larger decision space.

\section{Conclusion}~\label{sec:conclusion}

We presented \thesystem, a framework for cost-efficiently meeting latency SLOs on a multi-cluster cloud data warehouse, comprised of a proactive \component{Policy Tuner} that \textbf{plans} infrastructure scaling, a reactive \component{Autoscaler} that \textbf{adjusts} resources based on workload fluctuations and an online \component{Query Router} that \textbf{reacts} to concurrent load.
We showed that these components can work together to successfully meet latency SLOs of varying strictness, reducing cost by a mean of $26.4\%$ compared to the per-scenario next-best baseline on realistic Redbench workloads. Component-level evaluations showed that the \component{Query Router}
and \component{Autoscaler} respectively reduce SLO violation
rates by a mean of $47.8\%$ and $93.7\%$, relative to their corresponding alternatives.
Finally, we showed that the \component{Policy Tuner} can reduce the SLO violation rate by a mean of $44.6\%$ using a single day of workload history, and that each component is efficient given its intended operating timescale.

\begin{acks}
    This research was supported by Amazon, Google, and Intel as part
    of the MIT Data Systems and AI Lab (DSAIL). This research was also sponsored by the Department of the Air Force Artificial Intelligence Accelerator and was accomplished under Cooperative Agreement Number FA8750-19-2-1000. The views and conclusions contained in this document are those of the authors and should not be interpreted as representing the official policies, either expressed or implied, of the Department of the Air Force or the U.S. Government. The U.S. Government is authorized to reproduce and distribute reprints for Government purposes notwithstanding any copyright notation herein. 
\end{acks}

\bibliographystyle{ACM-Reference-Format}
\bibliography{autoslo}

\end{document}